\documentclass[reprint,twocolumn,prl,showpacs, amsmath, amssymb, superscriptaddress,aps]{revtex4-2}

\usepackage{graphicx}
\usepackage{dcolumn}
\usepackage{bm}
\usepackage{bbold}
\usepackage{amsmath}
\usepackage{comment}
\usepackage{xcolor}

\newcommand{\id}{\mathbb{1}} 

\begin{document}

\preprint{APS/123-QED}

\title{Mean path length invariance in wave-scattering beyond the diffusive regime}
\author{Matthieu Davy}
\affiliation{Univ.\ Rennes, CNRS, IETR (Institut d'{\'E}lectronique et des Technologies du num{\'e}Rique), UMR–6164, F-35000 Rennes, France}
\author{Matthias K\"uhmayer}
\affiliation{ Institute for Theoretical Physics, Vienna University of Technology (TU Wien), A-1040 Vienna, Austria}
\author{Sylvain Gigan}
\affiliation{ Laboratoire Kastler Brossel, Universit{\'e} Pierre et Marie Curie, {\'E}cole Normale Sup{\'e}rieure, CNRS, Coll\`{e}ge de France, F-75005 Paris, France}
\author{Stefan Rotter}
\affiliation{ Institute for Theoretical Physics, Vienna University of Technology (TU Wien), A-1040 Vienna, Austria}
\date{\today}

\begin{abstract}

Diffusive random walks feature the surprising property that the average length of all possible random trajectories that enter and exit a finite domain is determined solely by the domain boundary. Changes in the diffusion constant or the mean-free path, that characterize the diffusion process, leave the mean path length unchanged. Here we demonstrate experimentally that this result can be transferred to the scattering of waves, even when wave interference leads to marked deviations from a diffusion process. Using a versatile microwave setup, we establish the mean path length invariance for the crossover to Anderson localization and for the case of a band gap in a photonic crystal. We obtain these results on the mean path length solely based on a transmission matrix measurement through a novel procedure that turns out to be more robust to absorption and incomplete measurement in the localized regime as compared to an assessment based on the full scattering matrix. 

\end{abstract}


\begin{titlepage}
  \maketitle
\end{titlepage}

A fundamental result of diffusion theory is that the mean path length of particles diffusing across a certain region of space is entirely independent of the characteristics of the diffusion process \cite{Dirac1943,Case1967,Blanco2003}. In fact, only the shape of the outer boundary of this region determines the average length of particle trajectories between entering and exiting this region; whether the paths taken by the particles are straight lines or convoluted random walks is, however, completely irrelevant. As such, this mean path length invariance generalizes the so-called “mean chord length theorem” valid in the ballistic limit \cite{Case1967} and encompasses applications in basically all research fields where diffusion processes or random walks arise –- ranging from nuclear physics \cite{Dirac1943} and solar energy harvesting \cite{Vasiliev2019} to the movement of bacteria \cite{Frangipane2019}.  

The scope of this invariance property was recently expanded even further, when it was shown that not only particles, but also waves that scatter through a region of space are subject to the same invariance property \cite{Pierrat2014}. The key insight here is that the average ``time-delay'' associated with a scattering process is directly linked to the density of states (DOS) inside the scattering region \cite{Schwinger1951,Krein1962,Birman1992,Iannaccone1995}. From the Weyl law \cite{Weyl1911, Arendt} we then know that the DOS remains invariant when transforming a homogeneous medium in the ballistic limit into a diffusively scattering disordered medium of the same size \cite{Pierrat2014}. A recent experimental implementation followed exactly this line of thought by changing the turbidity of a liquid from nearly transparent to very opaque and demonstrated that the mean path length of isotropically incoming light inside the liquid stays, indeed, unchanged over nearly two orders of magnitude in scattering strength \cite{Savo2017}. In spite of its coherent nature, laser light could thus be observed to obey the same universal invariance property as particles when scattering ballistically or diffusively.

Coherent wave effects can, however, also lead to very strong deviations from any of the transport regimes that particles can be in. Consider here, e.g., the regime of Anderson localization \cite{Mirlin2000,Lagendijk2009} or the formation of a band gap in a photonic crystal \cite{Yablonovitch1993}, to cite just two genuinely wave-like phenomena that both rely on wave interference. The natural question to ask at this point is whether any such effects going beyond a trajectory-based description may lead to a violation of the mean path length invariance since they clearly fall outside the scope of both the mean chord length theorem and a random walk picture. More specifically, since both Anderson localization and a band gap prevent incident waves from propagating inside the scattering region, one naturally expects that the mean path length invariance should break down in these cases. Insights into such questions are not just of academic interest: consider, e.g., that the mean path length invariance is strongly linked to the so-called “Yablonovitch limit” that imposes a cap on the optimal intensity enhancement inside solar cells \cite{Yablonovitch1982}. A break-down of the invariance property may thus also provide a strategy for overcoming current limitations in solar cell design \cite{Vasiliev2019}.

\begin{figure*}
\includegraphics[width=16.5cm]{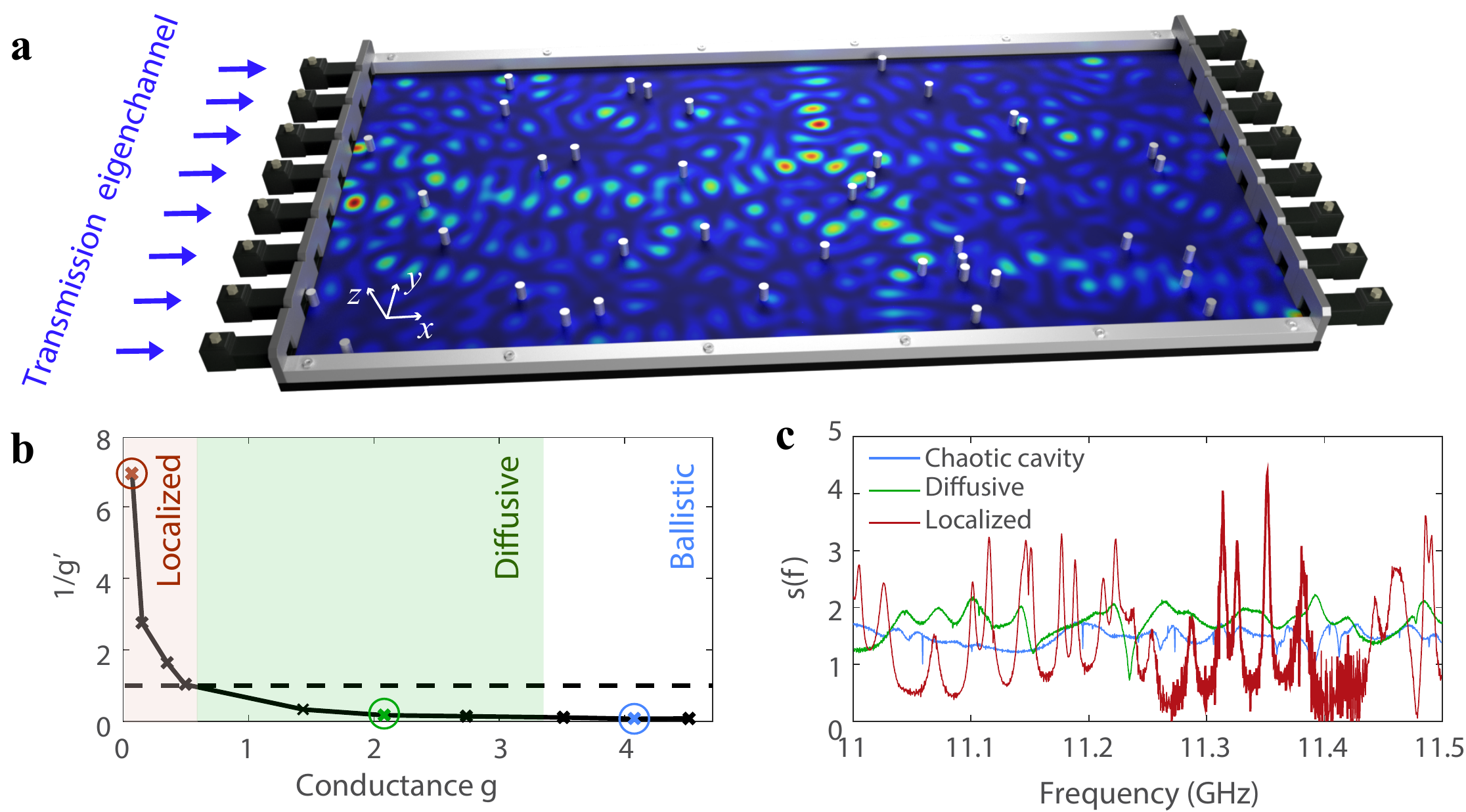}
\caption{\label{fig1} \textbf{Experimental setup and path length spectra.} \textbf{a}, Sketch of the experimental setup, where the cavity is represented for a random arrangement of 40 aluminum cylinders. The top plate has been removed to visualize the interior of the sample. The embedded intensity distribution corresponds to the highest transmitting transmission eigenstate in this scattering configuration obtained from a numerical simulation. \textbf{b}, Experimental data for the variation of the inverse of the statistical conductance, $(1/g')=3 \mathrm{var}(s_a)/2$ as a function of the conductance $g$ to estimate whether the experiment operates in the diffusive $[(1/g')<1]$ or in the localized regime $[(1/g')>1]$ \cite{Chabanov2000}. \textbf{c}, Spectra of the mean path length $s(\omega) = s[2Q_t(\omega)]$ for a chaotic cavity (blue line), a diffusive sample with $n_{\text{scat}}=100$ scatterers (green line) and a localized sample with $n_{\text{scat}}=280$ scatterers (red line). The values of $1/g'$ corresponding to these samples are shown with red, green and blue markers in \textbf{b}.}
\end{figure*} 

Here we will address these questions through an experiment that gives us direct access to these scattering regimes beyond both the ballistic and the diffusive limit. We test the invariance of the mean time-delay in microwave measurements using a multichannel cavity for which the scattering strength of a random sample can be tuned by changing the number of scatterers within the cavity and for which an ordered arrangement of the scatterers mimics a photonic crystal. 

The scattering region is formed by an effectively two-dimensional cavity (see Methods) of length $L = 0.5$ m and width $W = 0.25$ m with two arrays of $N = 8$ antennas attached on the left and right interfaces, respectively \cite{Davy2018}. The antennas are single mode waveguides fully coupled to the cavity between 11 and 18~GHz (see Fig.~\ref{fig1}a). Having full control over all these waveguides enables the measurement of the frequency-dependent $N\times N$ transmission matrix (TM) $t_{ba}(\omega)$, which contains the complex and flux-normalized transmission coefficients between the two arrays. The TM is complete, but strong internal reflections occur at the left and right interfaces of the cavity as the spacing between two adjacent antennas is metallic (unlike in open waveguides).

To induce a transition from the ballistic to the chaotic, diffusive and the localized regime we gradually increase the scattering strength within the cavity. Specifically, we measure the transmission matrix $t(\omega)$ for an empty cavity (ballistic system), for a cavity with an aluminum semicircle of 53 mm diameter [chaotic system, see Supplementary Material (SM) for details] and for disordered samples with $n_\text{scat}$ randomly distributed aluminum cylinders of radius $r_s=3$ mm. For the disordered systems the conductance $g=\langle \Sigma_{n=1}^{N} \tau_n \rangle$ (i.e., the frequency-averaged sum of transmission eigenvalues $\tau_n$ of $t^\dagger t$) ranges over more than two orders of magnitude from $4.36$ for the empty cavity to $0.01$ for the sample with the strongest disorder ($n_\text{scat} = 280$). 

 Because of dissipation within the system and strong internal reflections, the scaling of the conductance $g$ may not reflect the crossover from diffusive to localized waves found at $g = 1$ in open random systems \cite{Abrahams1979}. We exploit instead the statistics of the transmitted intensity as a reliable indicator of the localization transition even in the presence of absorption \cite{Chabanov2000}. Specifically, the variance of the normalized total transmission $s_a = T_a/\langle T_a \rangle$, with $T_a=\Sigma_b |t_{ba}|^2$, is given for non-dissipative diffusive samples by $\mathrm{var}(s_a) = 2/(3g)$. In the case of finite dissipation the so-called statistical conductance $g' = 2/[3\mathrm{var}(s_a)]$ has been found to indicate localization reliably when taking on values $(1/g')>1$ \cite{Chabanov2000}. In Fig.~\ref{fig1}b, we thus show the change of $1/g'$ with the conductance $g$ and observe values for $(1/g')$ that considerably exceed 1 for $g < 0.6$. This confirms that the increased disorder in our two-dimensional system leads to wave localization.

In multichannel systems, the mean path length can be estimated from measurements of the Wigner-Smith (WS) time-delay operator $Q=-iS^{-1}\partial_\omega S$ applied to the scattering matrix $S$ which relates incoming and outgoing channels \cite{Wigner1955,Smith1960,Kottos2005,Rotter2011,Gerardin2016,Boehm2018,Xiong2016,Ambichl2017,Carpenter2015}. The operator $Q$ is a multichannel generalization of the phase derivative $d\phi_{ba}/d\omega$, which provides the time-delay of a spectrally narrow pulse between two channels $a,b$. Averaging over all channels leads to the mean Wigner-Smith time-delay given by $\bar{t}_\textrm{WS}(\omega)=\mathrm{Tr}[Q(\omega)]/(2N)$.

In principle, estimating the mean time-delay requires a measurement of the complete scattering matrix $S(\omega)$ including the two reflection matrices on the left and right sides of the sample, respectively. Experimentally, such a measurement is highly challenging, however, since most setups provide access either only to a one-sided reflection matrix or to the TM. To overcome this difficulty, we show in the SM that the trace of $Q(\omega)$ and the trace of the Wigner-Smith operator involving only the TM from left to right, $Q_t(\omega)=-it^{-1} \partial_\omega t$, are connected through the following equivalence relation, 
\begin{equation}
s[Q(\omega)]= s[2 Q_t(\omega)] \ ,
\label{eq:trace}
\end{equation}
where $s(\mathcal{O}) = c_0 \mathrm{Re}[\mathrm{Tr}(\mathcal{O})/(2N)]$ is the mean length obtained with an operator $\mathcal{O}$. The above simple relation is proven in the SM for non-absorbing systems using the decomposition of the TM into transmission eigenchannels. The key ingredient is the correspondence between transmission and reflection eigenchannels as a consequence of the unitarity of the scattering matrix, $S(\omega) S^\dagger(\omega) = \id$ \cite{Brandbyge1998,Davy2015}. The transmission eigenchannel time-delay  $t_n^{(t)}(\omega) = d\theta_n/d\omega$, found from the derivative of a composite phase shift $\theta_n$ of the singular vectors of the TM \cite{Davy2015}, is then equal to the average of the corresponding reflection delay times at the right and left sides of the sample, $t_n^{(t)}(\omega) = [t_n^{(r)}(\omega)+t_n^{(r^\prime)}(\omega)]/2$ (as in 1D systems \cite{Avishai1985}), where a prime denotes the quantities at the other waveguide port. Equation \eqref{eq:trace} thus provides access to the mean path length $s[Q(\omega)] = c_0 \bar{t}_\textrm{WS} (\omega)$ through transmission measurements only.

\begin{figure}
\includegraphics[width=8.5cm]{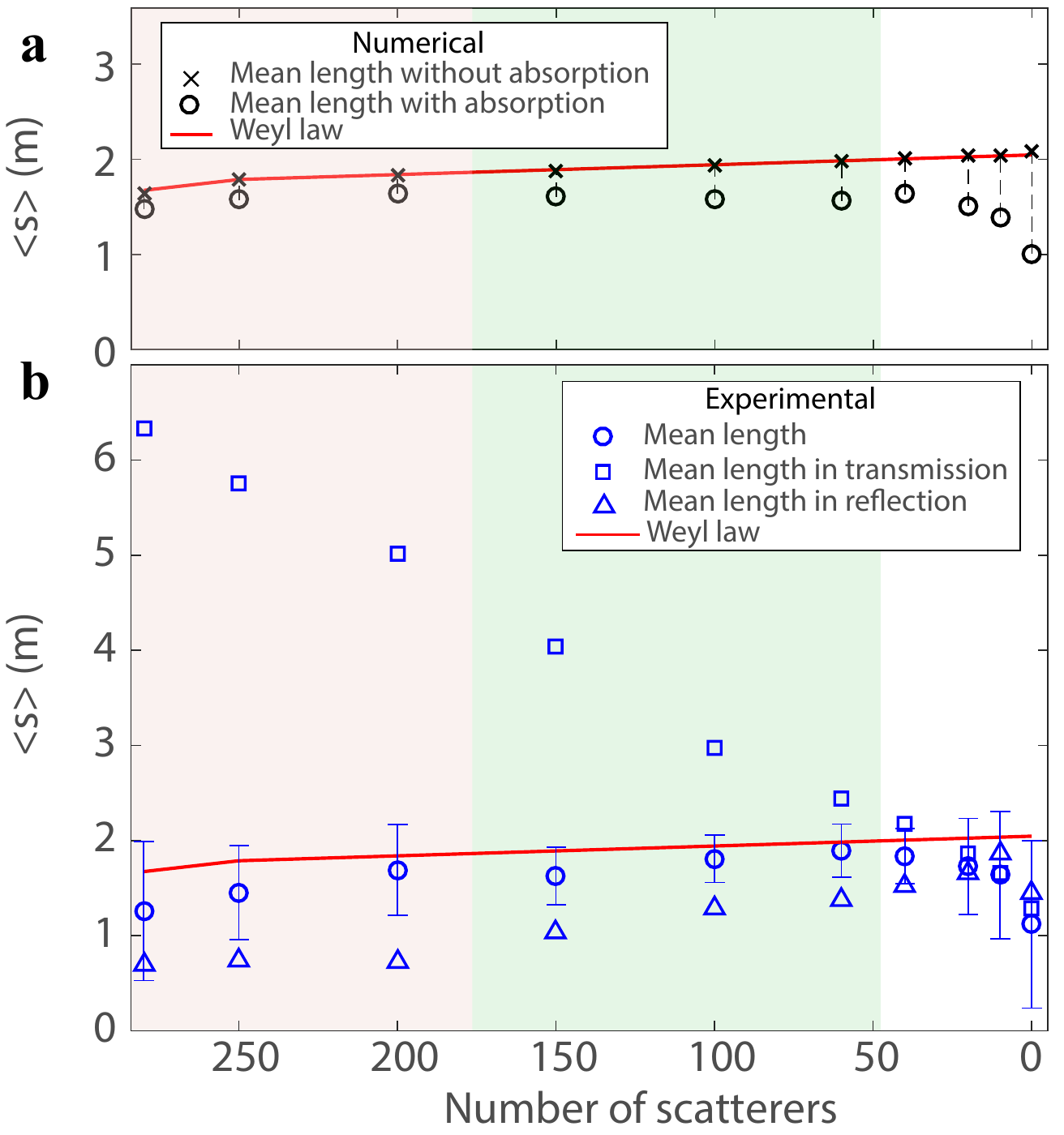}
\caption{\label{fig2} \textbf{Mean path lengths with respect to the number of scatterers.} \textbf{a}, Mean path length evaluated as $\langle s(2Q_t) \rangle$, see Eq. (\ref{eq:trace}), obtained in simulations without (black crosses) and with (black circles) absorption as a function of the number of metallic cylinders within the cavity. The red solid line is the theoretical length calculated from the Weyl law [see Eq.~\eqref{eq:mean_length}]. Absorption reduces the estimated mean path length. Background colors in rose, green, and white mark the localized, diffusive, and the ballistic regime as in Fig.~\ref{fig1}b. \textbf{b}, The same quantity, $\langle s(2Q_t) \rangle$, found experimentally (blue circles with error bars). The squares and triangles are the mean path length in transmission and in reflection, respectively (see SM for details). }
\end{figure}

In practice, evaluating $s[2Q_t(\omega)]$ requires that eigenchannels of $t(\omega)$ with small transmission eigenvalues $\tau_n$ (typically $\tau_n < 10^{-6}$) are removed from the experimental data since these eigenchannels may contain time-delays $t_n^{(t)}(\omega)$ that are corrupted by the noise level of the experimental setup. In diffusive and localized systems, eigenchannels with the smallest transmission values are typically associated with a small intensity build-up inside the medium and therefore with small time-delays \cite{Davy2015}. Removing their contribution to $s[2Q_t(\omega)]$ thus modifies the estimated length only weakly (see SM).

Spectra of the mean path length $s[2Q_t(\omega)]$ corresponding to a chaotic cavity, a diffusive sample and a localized sample, are presented in Fig.~\ref{fig1}c. In contrast to the diffusive regime in which the overlap of many resonances leads to small fluctuations of $s[2Q_t(\omega)]$, the peaks observed in the localized regime correspond to the contributions of spectrally isolated resonances associated with localized modes. To meaningfully compare the mean path length obtained in the different propagation regimes, it is thus necessary to average $s[2Q_t(\omega)]$ over a frequency range containing several of these peaks. 

In Fig.~\ref{fig2} we compare the mean path length $\langle s(2Q_t) \rangle$ resulting from a frequency-average in single configurations (see Methods) with the theoretical predictions obtained using the Weyl law \cite{Weyl1911,Arendt,Pierrat2014} given by, 
\begin{equation}
s_\mathrm{theo}(\omega) = \frac{1}{2Nc_0} \left( \omega A+\frac{C-B}{2}c_0 \right)\,.
\label{eq:mean_length}
\end{equation}

\noindent This Weyl prediction for the mean path length involves the scattering area $A = L W + 16 l w - n_{\text{scat}}\pi r_s^2$ corresponding here to the surface of the cavity, including the attached channels of length $l = 38$ mm and aperture $w = 15.79$ mm, from which the area of the impenetrable metallic scatterers is subtracted. A first-order correction term takes into account the external boundaries $C = 16 w$ and the internal boundaries $B$ of the scattering system, that include the metallic boundaries of the cavity and the circumferences of the metallic cylinders \cite{Pierrat2014}. Smaller values of $g$ are reached with an increasing number of these cylinders, which reduces the effective scattering area and increases the length of the internal boundaries. The theoretical estimate of the mean path length $\langle s_\mathrm{theo}(\omega) \rangle$ is then the average of $s_\mathrm{theo}(\omega)$ over the frequency range and decreases from 2.05~m for an empty cavity to 1.76~m for a sample with 280 cylinders. The theoretical mean path length for an empty cavity $\langle s_\mathrm{theo}(\omega) \rangle \sim 4L$  strongly exceeds its value for open waveguides $\langle s_\mathrm{theo}(\omega) \rangle =  \pi L /2$ \cite{Pierrat2014}. This enhancement is due to the metallic spacings between the antennas at the right and left interfaces of the cavity. Numerical simulations presented in the SM also show the existence of states with very long delay times in the empty cavity, corresponding to path lengths of a few hundred meters. These states are caused by bouncing orbits between the top and bottom interfaces of the cavity (in $y$-direction).

Figure~\ref{fig2}a shows perfect agreement between the Weyl prediction (see red line) and the mean path length $\langle s(2Q_t) \rangle$ obtained in numerical simulations (see black crosses) of the experimental setup in absence of absorption (see Methods). Even the small reduction in the mean path length predicted for an increasing number of scatterers is well reproduced. This confirms the validity of the mean path length invariance across the onset of the localization transition.

Also the experimental data shown in Fig.~\ref{fig2}b are in good agreement with $\langle s_\mathrm{theo}(\omega) \rangle$ (see blue circles). How non-trivial this invariance property is, can be appreciated when contrasting it with the strong enhancement (reduction) of the transmission (reflection) time-delays across the localization transition (see blue squares and triangles). We also observe that the mean path length is slightly underestimated, especially in the ballistic regime. Even though the DOS integrated over frequency is independent of absorption \cite{Barnett1996}, the presence of absorption within the cavity makes the scattering matrix sub-unitary \cite{Fyodorov2005} and leads to a violation of the mean path length invariance \cite{Pierrat2014} so that this deviation comes as no surprise. Indeed, the experimental data is well reproduced by numerical simulations when dissipation is included. For the data shown in Fig.~\ref{fig2}a (see black circles), we introduce uniform absorption by adding an imaginary part to the effective refractive index of the cavity. We here use for the empty cavity as well as for all disorder configurations an average uniform imaginary part of the refractive index  ($n_i = 2 \times 10^{-4}$) found by comparing the frequency-averaged transmission measured for the empty cavity with the numerical simulations.

In addition to its detrimental effect on the mean path length invariance, dissipation also breaks the unitarity of $S$ and therefore the correspondence between $s[Q(\omega)]$ and $s[2Q_t(\omega)]$ in Eq.~(\ref{eq:trace}). In the diffusive regime, all of these effects are sufficiently weak such that losses through outgoing channels dominate over uniform absorption over the sample. From the broadening of the average linewidth with respect to the number of ports connected to the cavity, we estimate that the ratio of losses through ports relative to uniform losses within the cavity is equal to 5.4 (see SM). With the scattering matrix then still being sufficiently  close to unitarity, $\langle s(2Q_t) \rangle$  provides a reliable estimator of the theoretical value $\langle s_\text{theo} \rangle$ in absence of absorption. Interestingly, the stronger deviations found for the empty cavity are a consequence of very specific states that bounce many times between the top and bottom cavity boundary (in $y$-direction). Due to their long cavity dwell times, these states are very strongly affected by dissipation and therefore lead to significant deviations from the mean path length invariance (see SM). The existence of these states is reflected by the large fluctuations of the mean path length in Fig.~\ref{fig2}. In the diffusive regime, the disorder scattering naturally leads to a suppression of such states with strongly enhanced time-delays and thereby to a better agreement with the theoretical predictions.

\begin{figure}
\centering
\includegraphics[width=8.5cm]{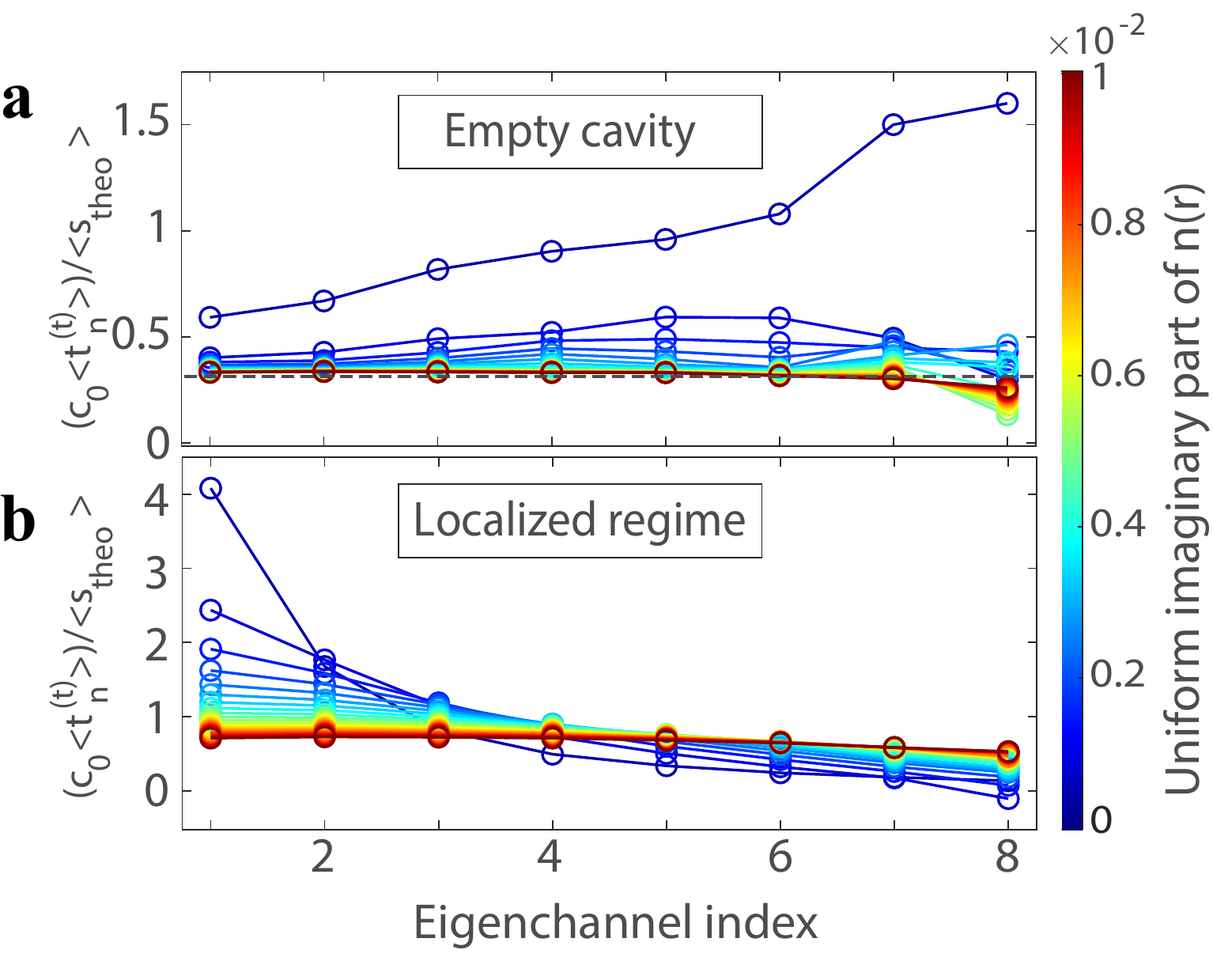}
\caption{\textbf{Influence of absorption.} \textbf{a},\textbf{b}, Normalized transmission eigenchannel contribution to the mean path length as a function of the index $n$, where $n=1$ ($n=8$) corresponds to the highest (lowest) transmitting eigenchannel. Increasing absorption leads to a redistribution of the transmission time-delays which all converge to the length of the most direct path in the limit of strong absorption (see SM). The horizontal black dashed line marks an estimation of the most direct path in the empty system.
}
\label{fig:mean_length_absorption}
\end{figure}

Surprisingly, we observe that our estimate for the mean path length, $\langle s(2Q_t) \rangle$, is more robust to absorption in the localized regime than in the empty cavity even though Anderson localization also gives rise to the large time-delays in the cavity and large fluctuations of the mean path length (see Fig.~\ref{fig2}b). Using a toy model, we demonstrate analytically in the SM that in the limit of strong absorption $s[2Q_t(\omega)]$ converges towards the most direct path in transmission $ \langle L_{\text{direct}} \rangle$. As shown in Fig.~\ref{fig:mean_length_absorption}, all eigenchannels feature a quasi-ballistic propagation in this limit providing the same contribution to the mean path length. In this way, the flux in the empty system (with $n_i = 10^{-2}$, see SM) becomes almost perfectly aligned with the $x$-direction from left to right as strong absorption suppresses all longer paths \cite{Liew2014}. Indeed, the resulting estimate for the most direct path $\langle L_\mathrm{direct}^\mathrm{empty} \rangle \approx 0.31 \langle s_\mathrm{theo} \rangle$ is in very good agreement with the numerical value of $\langle s(2Q_t) \rangle = 0.32 \langle s_\mathrm{theo} \rangle$. In the localized regime, the presence of metallic and hence impenetrable scatterers elongates the direct path, which is found to be comparable to the mean path length ($\langle s(2Q_t) \rangle = 0.66 \langle s_\mathrm{theo} \rangle$ for the sample with 280 scatterers and $n_i = 10^{-2}$). The increased absorption-stability of the mean path length in the localized regime as compared to the ballistic regime is thus explained by the difference in the most direct scattering contributions in these two cases.
We emphasize that this robustness of $s[2Q_t(\omega)]$ in the localized regime is in stark contrast to the much stronger dependence on absorption we observe for a corresponding estimate using the full Wigner-Smith matrix $s[Q(\omega)]$ (see SM). $Q(\omega)$ contains the full scattering matrix and thus depends on the sum of phase delay times of both transmitted and reflected waves 
\cite{AmbichlPhd2016}. In analogy to our analysis from above, the strong absorption reduces these contributions to those coming from the shortest possible paths. However, since the shortest paths contributing here are those that are directly reflected when entering the cavity, their extremely short time-delay values will dominate in the strong absorption limit (see SM). Compared to transmission eigenchannels whose direct paths have to traverse the whole system, the convergence of $s[Q(\omega)]$ to the direct paths in reflection leads eventually to a pronounced underestimation of the mean path length for strongly scattering samples.

\begin{figure}
\includegraphics[width=8.5cm]{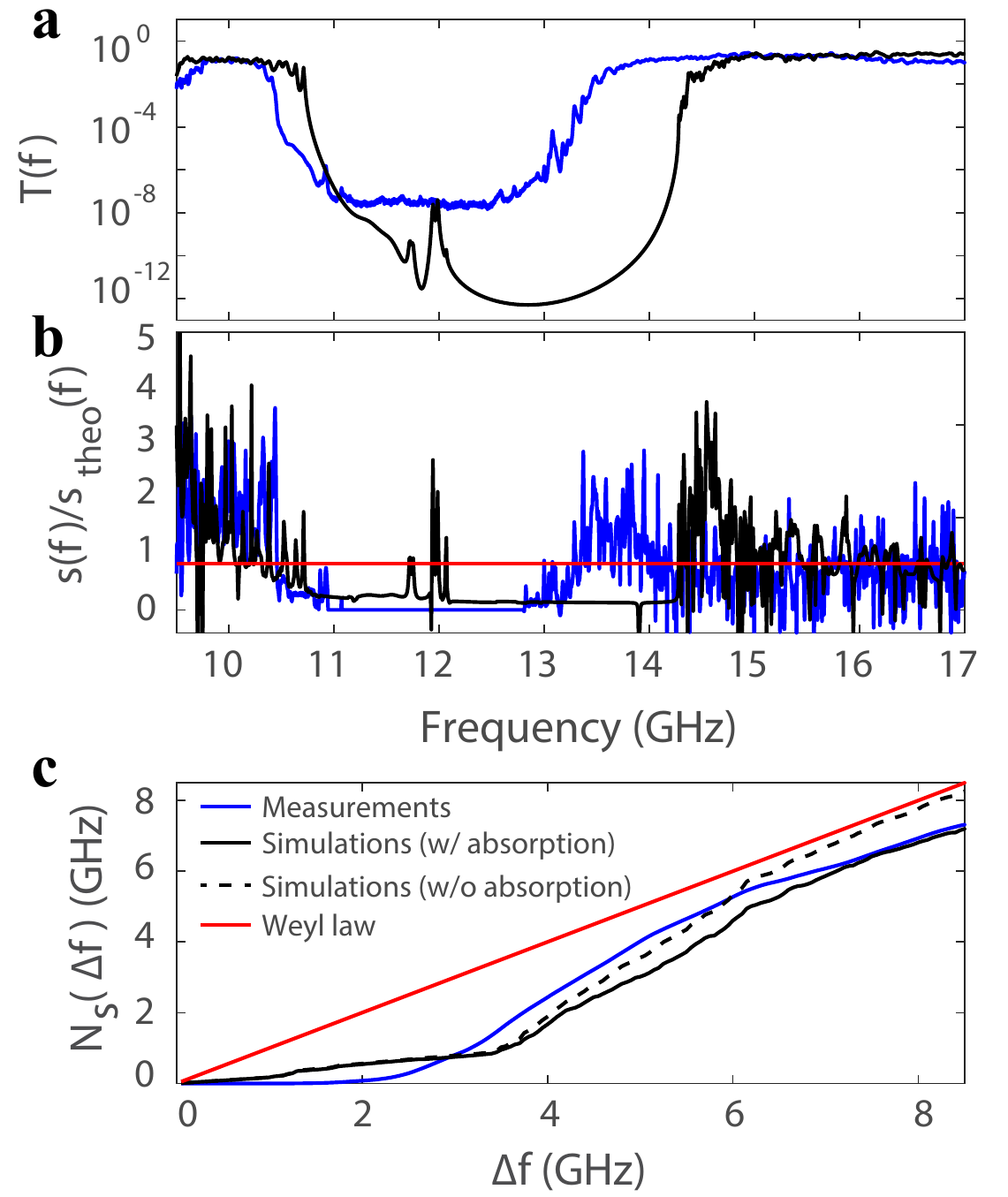}
\caption{\textbf{Path lengths in a photonic crystal.} \textbf{a},\textbf{b}, Frequency spectra \textbf{a} of the transmission and \textbf{b} of the mean path length in a photonic crystal obtained in measurements to which the free space contribution has been removed outside the band gap (blue line) and in simulations with absorption (black line). A transmission band gap is clearly observed for this sample with 15 longitudinal layers of regularly spaced scatterers. \textbf{c}, Integral of the mean path length normalized by its theoretical value over a frequency window spanning $\Delta f$ around the center of the bandgap. The black dashed line is the result of simulations in absence of absorption. The decrease of the mean path length within the band gap is followed by a strong enhancement starting at the band edges such that the frequency-averaged value progressively converges towards the theoretical Weyl prediction. The experimental data (blue solid line) shows the same trend, but stays slightly below the Weyl prediction for large $\Delta f$ due to absorption (see agreement with the simulations including absorption, black solid line).}
\label{fig4}
\end{figure}

After having explored the validity of the mean path length invariance in the cross-over to very strong disorder, we will now consider the opposite limit of a structured medium with periodic order. For this purpose we form a photonic crystal (PC) by a periodic arrangement of alternating aluminum and Teflon cylinders (see Methods). The structure with 15 layers of scatterers in logitudinal direction exhibits a band gap centered at $f_0 = 12
~\mathrm{GHz}$ with a width $\Delta f_0 \approx 2~\mathrm{GHz}$ as can be seen in Fig.~\ref{fig4}a. Because the PC only fills the middle part of the entire scattering area of the cavity, the mean path length includes the contribution of the PC as well as the free space between the PC and the interfaces of the cavity. To isolate the impact of the PC, we subtract the theoretical free space contribution to $s[2Q_t(\omega)]$, except in the band gap for which the transmission remains in the noise floor of the experimental setup ($\approx 10^{-6}$) and we set $s[2Q_t(\omega)] = 0$.

Numerical simulations are then performed in a cavity whose scattering region has the same dimensions as the PC with small disorder in the scatterer positions and absorption being added to mimic the situation in the experiment (see Methods). The central frequency of the band gap is now $f_0 = 12.6~\mathrm{GHz}$ with a width of 3.3~GHz, which is larger than in measurements. We attribute these differences to tiny air gaps between the top plate and the aluminum cylinders in the experimental setup, where scattering at the top cylinder edge causes the excitation of evanescent modes. The effective properties of the     scatterers are thus modified locally due to the coupling of such evanescent modes to neighboring Teflon scatterers. 
The mean path length is also strongly reduced within the band gap in simulations (see Fig.~\ref{fig4}b), but $s[2Q_t(\omega)]$ is not vanishing as a consequence of the finite length of the PC. We also note that the dwell time can be negative as a result of absorption \cite{Durand2019} and since the dwell time operator is related to the time-delay operator and thus also to $s[2Q_t(\omega)]$, this explains the negative values in Fig.~\ref{fig4}b.

From such an observation one may be tempted to conclude that the mean path length invariance does not hold in such periodic systems. Following prior theoretical work on frequency sum rules \cite{Barnett1996,Carminati2009}, we know, however, that in a sufficiently broad spectral window, reductions and enhancements of the DOS should compensate each other in arbitrary systems including the case of photonic bandgap materials. Due to the connection between the DOS and the mean path length, we should find, correspondingly, that the strong decrease of $s[2Q_t(\omega)]$ within the band gap is compensated by a corresponding enhancement right outside of the band gap. As can be seen in Fig.~\ref{fig4}b, we indeed observe such an enhancement of the mean path length close to the band edges, with values for $s[2Q_t(\omega)]$ even exceeding $3 s_\text{theo}(\omega)$ in measurements and in simulations. 

In Fig.~\ref{fig4}c we now show the mean path length normalized by Weyl prediction, $s[2Q_t(f)]/s_\text{theo}(f)$, integrated over a frequency window $\Delta f$ centered around the middle of the corresponding band gap $f_0$ obtained in the experiment or the simulation, $N_s(\Delta f) = \int_{f0-\frac{\Delta f}{2}}^{f0+\frac{\Delta f}{2}}\,  s[2Q_t(f)]/s_{\textrm{theo}}(f) \, df$. The Weyl law prediction for this quantity is $N_s(\Delta f)=\Delta f$. Note that the bandgap is not located in the center of our frequency range. Thus, once the lower end of the integration frequency window has reached 9.5~GHz, we continue the integration with only the higher frequency range. 
In the experiment, this average ratio $N_s(\Delta f)$ now almost vanishes for a spectral window $\Delta f$ smaller than the width of the band gap but then increases rapidly and progressively converges towards $N_s(\Delta f) \approx 0.86 \Delta f$ as $\Delta f = 8.5~\mathrm{GHz}$. As in the case of the empty system or the disordered configurations, this 14\% deviation from the Weyl law is due to absorption within the sample. This is confirmed by simulations without absorption for which $N_s(\Delta f)$ indeed reaches $N_s(\Delta f) \approx 0.975 \Delta f$ . The convergence of $N_s(\Delta f)$ now demonstrates that the pronounced enhancements of $s[2Q_t(\omega)]$ at the edges of the band gap compensate the vanishing mean path length within the band gap. The fact that we observe the experimental data and the simulations with absorption to converge towards the same value of $N_s$ for large $\Delta f$ in spite of the different band gap sizes for these two cases, further substantiates the invariance property of the mean path length. 

In summary, we experimentally demonstrated with microwave measurements that the knowledge of the transmission matrix alone provides a robust way of estimating the invariant mean path length, even in scattering systems with  strong wave-interference and weak absorption. Our results clearly show that this invariance property reaches far beyond the diffusive regime and thus provides a comprehensive bound on enhancement of the mean path length of broadband light in a medium.

\section{Methods}

\noindent \textbf{Experimental setup.} Sixteen antennas that are waveguide-to-coax adapters operating in the Ku-band are attached to the system and fully coupled to the cavity between 11 and 18~GHz. We operate in a frequency window smaller than $c_0/2h = 18.75~\mathrm{GHz}$ where only the fundamental mode in the vertical $z$-direction is excited, making the cavity effectively two-dimensional. Measurements of the frequency spectra of the TM are carried out with two ports of a vector network analyzer connected to two $N \times 1$ electro-mechanical switches to successively excite each transmitting and receiving antenna. The ports of the switches that are not excited are terminated with 50~$\Omega$ loads so that the antennas mimic absorbing boundary conditions. The metallic spacing between adjacent antennas generate strong internal reflections at the interfaces of the cavity.
The TM is measured between 11 and 13~GHz in the ballistic and diffusive regimes and between 11 and 12~GHz in the localized regime, with frequency steps of 0.4~MHz. \\

\noindent \textbf{Photonic crystal.} The period of the square lattice between a metallic and a dielectric Teflon scatterer is $d = 1.2$ cm in both the longitudinal and transverse direction. A picture of the experimental setup is shown in the SM. The photonic crystal consists of 15 longitudinal layers of scatterers. The variations of the theoretical mean path length $s_\text{theo}(\omega)$ with frequency for a sample of 300 cylinders is given by Eq. (2), where the area of the Teflon cylinders has to be multiplied with their refractive index squared to account for the increased density of states in dielectric materials. \\

\noindent \textbf{Numerical simulations.} We solve the two-dimensional scalar Helmholtz equation $\left[ \Delta + n^2(\mathbf{r}) k_0^2 \right] \psi (\mathbf{r}) = 0$ using a finite element method \cite{NGSolve, Schoberl1997,Schoberl2014}. Here, $\Delta$ is the Laplacian in two dimensions, $n(\mathbf{r})$ is the refractive index distribution, $\mathbf{r} = (x,y)$ is the position vector, $k_0 = 2 \pi/\lambda$ is the vaccum wavenumber and $\psi (\mathbf{r})$ is the $z$-component of the TE-polarized electric field. In our simulations we use the exact dimensions of the experimental setup (see Fig.~\ref{fig1}), where the single mode leads are terminated with perfectly matched layers which absorb the outgoing waves without any back-reflections and thus mimic semi-infinite leads. To account for the global losses in the experiment, we add a frequency-independent uniform imaginary part of $2 \times 10^{-4}$ to the effective refractive index of the cavity which yields the same frequency-averaged transmission in the empty system as in the experiment.

In case of the disordered systems, we use -- just like in the experiment -- a single random configuration for each number of scatterers and average the calculated mean path lengths in the range of 11-13~GHz (11-12~GHz for 280 scatterers).

The simulations of the photonic crystal are performed in the frequency interval of 9.5-17~GHz. To mimic experimental uncertainties, we introduce a slight disorder to the positions of the scatterers, i.e., we displace scatterers in the transverse and longitudinal direction by a random value drawn from the interval $[-r_s/10, r_s/10]$. We then define our bandgap as the frequency interval in which the transmission in the numerical simulation reaches the experimental noise floor ($\approx 10^{-6}$).

To calculate the time-delay operators $Q$, $Q_t$, $Q_r$ and $Q_{r^\prime}$ we have to invert the corresponding scattering, transmission or reflection matrices. Since these matrices can be singular, we perform a singular value decomposition and project our matrices onto subspaces containing only singular vectors corresponding to singular values greater than $10^{-10}$ which enables us to compute their pseudo-inverse (see Supplementary Material of \cite{Ambichl2017} or \cite{Brandstoetter2019} for details).

\section{Acknowledgments}
\noindent This publication was supported by the European Union through the European Regional Development Fund (ERDF), by the French region of Brittany and Rennes M{\'e}tropole through the CPER Project SOPHIE/STIC \& Ondes, and by the Austrian Science Fund (FWF) through project P32300 (WAVELAND). The computational results presented were achieved using the Vienna Scientific Cluster (VSC). M. D. acknowledges the Institut Universitaire de France.

\section{Author contributions}
\noindent Measurements and data evaluation were carried out by M.D. Numerical simulations were carried out by M.K. under the supervision of S.R. Theoretical tasks were carried out by M.D., M.K., S.R. and S.G. M.D., M.K. and S.R. wrote the manuscript with input from all authors.

\section{Competing interests}
\noindent The authors declare no competing interests.

\section{Data availability}
\noindent The data that support the plots within this paper and other findings of this study are available from the corresponding authors on reasonable request.

\clearpage

\renewcommand{\thefigure}{S\arabic{figure}}
\renewcommand{\theequation}{S\arabic{equation}}
\setcounter{equation}{0}
\setcounter{figure}{0}
\setcounter{section}{0}

\date{\today}

\title{Supplementary Material for 'Mean path length invariance in wave-scattering beyond the diffusive regime'}
\begin{titlepage}
\maketitle
\end{titlepage}

\section{Measurements}
\subsection{Experimental Setup} 
\begin{figure}[!htb]
\includegraphics[width=8.5cm]{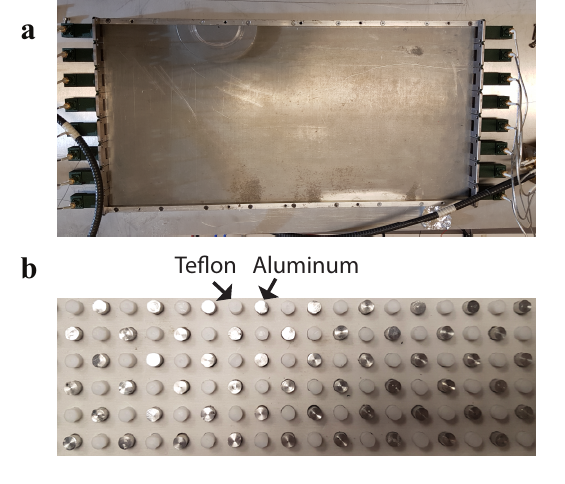}
\caption{\label{fig_cavity} \textbf{a}, Picture of the cavity with a semicircle. \textbf{b}, Picture of the ordered arrangement of aluminum and teflon cylinders.}
\end{figure}

Figure~\ref{fig_cavity}a shows a picture of the chaotic cavity. The cavity is made chaotic by including an aluminum semicircle of 53 mm diameter. The semicircle is located near a boundary.

We also present in Fig.~\ref{fig_cavity}b a picture of the the structured medium forming a photonic crystal. The periodic order is made by the arrangement of aluminum and teflon cylinders. The distance between two cylinders is equal to $d = 1.2$ cm in both the longitudinal and transverse direction. The number of layers of scatterers is equal to 15.

\subsection{Mean path lengths in transmission and reflection}
In Fig.~2 of the main text, we compare the mean path length with the mean path lengths in transmission $\langle s_t \rangle$ and in reflection $\langle s_r \rangle$. In transmission, $\langle s_t \rangle$ is found using the average of the single-channel delay times with 
\begin{equation}
\langle s_t \rangle = c_0 \frac{\langle T_{ba} \frac{d\phi_{ba}}{d\omega}\rangle}{\langle T_{ba}\rangle}.
\end{equation}
Here $T_{ba}$ and $\phi_{ba}$ are the transmitted intensity and the phase associated to the element $t_{ba}$ of the transmission matrix, i.e., the transmission coefficient between incoming channel $a$ and outgoing channel $b$, $t_{ba}=\sqrt{T_{ba}}\exp(i\phi_{ba})$. The averaging is performed over incoming and outgoing channels as well as the frequency range and $c_0$ is the speed of light.

In case of reflection, our experimental setup does not allow the measurement of the full reflection matrix. However, we obtain its diagonal part with elements $r_{aa} = \sqrt{R_{aa}} \exp(i\phi_{aa})$ using the $S_{11}$ parameters of the Vector Network Analyzer. The mean reflection length is then computed from
\begin{equation}
\langle s_r \rangle = c_0 \frac{\langle R_{aa} \frac{d\phi_{aa}}{d\omega}\rangle}{\langle R_{aa}\rangle}.
\end{equation}

\section{Derivation of Eq.~(1) of the main text}
In the following we give an analytical proof of Eq.~(1) of the main text which reduces to
\begin{equation}
\mathrm{Tr} (Q) = 2 \mathrm{Re} \left[ \mathrm{Tr} (Q_t) \right] \ ,
\end{equation}
with the time-delay operators 
\begin{align}
&Q = -i S^{-1} \frac{dS}{d\omega} \ , \\
&Q_t = -i t^{-1} \frac{dt}{d\omega} \ ,
\end{align}
in case of a two-port scattering matrix
\begin{equation}
S = \begin{pmatrix}
r & t' \\
t & r' \\
\end{pmatrix} ,
\end{equation}
which is unitary, i.e., $S^{-1} = S^\dagger$ and thus $S^\dagger S = \id$. Here, primed quantities denote the transmission or reflection from the other port. Thus, we can write the $Q$-operator as
\begin{equation}
Q =  \begin{pmatrix}
-i r^\dagger \frac{dr}{d\omega} -i t^\dagger \frac{dt}{d\omega} & -i r^\dagger \frac{dt'}{d\omega} -i t^\dagger \frac{dr'}{d\omega} \\
-i t^{\prime \, \dagger} \frac{dr}{d\omega} -i r^{\prime \, \dagger} \frac{dt}{d\omega} & -i r^{\prime \, \dagger} \frac{dr^{\prime \, \dagger}}{d\omega} -i t^{\prime \, \dagger} \frac{dt^{\prime \, \dagger}}{d\omega} 
\end{pmatrix} ,
\end{equation}
where its trace is given by
\begin{align}
\begin{split}
\mathrm{Tr} (Q) = \mathrm{Tr} \bigg( &-i t^\dagger \frac{dt}{d\omega} -i t^{\prime \, \dagger} \frac{dt^{\prime \, \dagger}}{d\omega} \\
&-i r^\dagger \frac{dr}{d\omega} -i r^{\prime \, \dagger} \frac{dr^{\prime \, \dagger}}{d\omega} \bigg) .
\end{split}
\label{eq:trQ_start}
\end{align}
To calculate the first term, we use a singular value decomposition (SVD) of the transmission matrix $t = U \Sigma V^\dagger$, where $U = (\vec{u}_1, \vec{u}_2, \ldots, \vec{u}_N)$ and $V = (\vec{v}_1, \vec{v}_2, \ldots, \vec{v}_N)$ are the matrices which contain column-wise the left and right singular vectors of $t$ and $\Sigma = \mathrm{diag\{ \sigma_n \}}$ is a diagonal matrix containing the singular values $\sigma_n$. Making use of $U^\dagger U = \id$ and $V^\dagger V = \id$ and the invariance of the trace with respect to cyclic permutations we can write
\begin{align}
\begin{split}
\mathrm{Tr} \left( -i t^\dagger \frac{dt}{d\omega} \right) = \mathrm{Tr} \bigg( 
&-i \Sigma^2 U^\dagger \frac{dU}{d\omega} \\
&-i \Sigma^2 \frac{dV^\dagger}{d\omega} V 
-i \Sigma \frac{d\Sigma}{d\omega} \bigg) .
\end{split}
\end{align}
Using $V^\dagger V = \id$, we can rewrite the second term in the trace as  $\frac{dV^\dagger}{d\omega} V = -V^\dagger \frac{dV}{d\omega}$.
Making further use of the fact that the singular values of $t$ are the square root of the eigenvalues of $t^\dagger t$ and $t t^\dagger$, i.e., $\sigma_n^2 = \tau_n$ are the transmission values, we arrive at
\begin{align}
\begin{split}
\mathrm{Tr} \left( -i t^\dagger \frac{dt}{d\omega} \right) = \sum_n \tau_n \ \frac{d\theta_n^{(t)}}{d\omega} - i \sigma_n^{(t)} \frac{d\sigma_n^{(t)}}{d\omega} \ .
\end{split}
\label{eq:trQt_dagger}
\end{align}
Here we have introduced the channel delay times as
\begin{equation}
\frac{d\theta_n^{(t)}}{d\omega} \equiv \frac{1}{i} \left( \vec{u}_n^{\, (t) \dagger} \frac{d\vec{u}_n^{\, (t)}}{d\omega} - \vec{v}_n^{\, (t) \dagger} \frac{d\vec{v}_n^{\, (t)}}{d\omega} \right)\ ,
\end{equation}
where the superscript denotes in the following to which matrix a singular vector (or singular value) belongs to.

To calculate the remaining terms, we make use of the relations between the singular vectors which follow directly from the unitarity of the scattering matrix. For the right singular vectors $\vec{v}_n$ we get
\begin{align}
t^\dagger t + r^\dagger r = \id \quad &\Rightarrow \quad \vec{v}_n^{\, (t)} = \vec{v}_n^{\, (r)} \equiv \vec{v}_n^{\, (t,r)} \ , 
\label{eq:right_sv1} \\
t^{\prime \, \dagger} t' + r^{\prime \, \dagger} r' = \id \quad &\Rightarrow \quad \vec{v}_n^{\, (t')} = \vec{v}_n^{\, (r')} \equiv \vec{v}_n^{\, (t',r')} \ ,
\label{eq:right_sv2}
\end{align}
The left singular vectors $\vec{u}_n$ are related as follows
\begin{align}
t t^\dagger + r' r^{\prime \, \dagger} = \id \quad &\Rightarrow \quad \vec{u}_n^{\, (t)} = \vec{u}_n^{\, (r')} \equiv \vec{u}_n^{\, (t,r')} \ , 
\label{eq:left_sv1} \\
t' t^{\prime \, \dagger} + r r^\dagger = \id \quad &\Rightarrow \quad \vec{u}_n^{\, (t')} = \vec{u}_n^{\, (r)} \equiv \vec{u}_n^{\, (t',r)} \ .
\label{eq:left_sv2}
\end{align}
Furthermore, we know that $t' = t^T$ which implies that $t$ and $t'$ share the same eigenvalues. Thus, $\sigma_n^{(t)} = \sigma_n^{(t')}$ and since ${\sigma_n^{(t)}}^2 = \tau_n$ it follows that ${\sigma_n^{(r)}}^2 = \rho_n = 1-\tau_n = 1-\tau'_n = \rho'_n = {\sigma_n^{(r')}}^2$. Using now Eq.~\eqref{eq:trQt_dagger} and similar expressions for the terms in Eq.~\eqref{eq:trQ_start} containing $t'$, $r$ and $r'$ yields
\begin{align}
\begin{split}
\mathrm{Tr} (Q) = \sum_n &\frac{1}{i} \Bigg[ 
\vec{u}_n^{\, (t,r') \dagger} \frac{d\vec{u}_n^{\, (t,r')}}{d\omega} - \vec{v}_n^{\, (t,r) \dagger} \frac{d\vec{v}_n^{\, (t,r)}}{d\omega} \\
&+ \vec{u}_n^{\, (t',r) \dagger} \frac{d\vec{u}_n^{\, (t',r)}}{d\omega} - \vec{v}_n^{\, (t',r') \dagger} \frac{d\vec{v}_n^{\, (t',r')}}{d\omega} \Bigg] \\
&-i \left[ \frac{d{\sigma_n^{(t,t')}}^2}{d\omega} + \frac{d{\sigma_n^{(r,r')}}^2}{d\omega} \right] ,
\end{split}
\label{eq:trQ_terms}
\end{align}
where we have already used the relations \eqref{eq:right_sv1}-\eqref{eq:left_sv2}. The last term corresponds now to the imaginary part of the $Q$-eigenvalues which has to vanish because $Q$ is a Hermitian operator (if $S$ is unitary) featuring real eigenvalues. Using $\tau_n^{(\prime)} + \rho_n^{(\prime)} = 1$ which follows from the unitarity of $S$ then indeed yields
\begin{equation}
\frac{d{\sigma_n^{(t,t')}}^2}{d\omega} + \frac{d{\sigma_n^{(r,r')}}^2}{d\omega} = \frac{d(\tau_n^{(\prime)} + \rho_n^{(\prime)})}{d\omega} = 0 \ .
\end{equation}
Eq. \eqref{eq:trQ_terms} can now be written in two different ways:
\begin{align}
\begin{split}
\mathrm{Tr} (Q) &= \sum_n \frac{d\theta_n^{(t)}}{d\omega} + \frac{d\theta_n^{(t')}}{d\omega} \\
&= \sum_n \frac{d\theta_n^{(r)}}{d\omega} + \frac{d\theta_n^{(r')}}{d\omega} \ .
\end{split}
\label{eq:trQ}
\end{align}
Next, we want to connect this result to the trace of $Q_t$, which can be simplified as follows:
\begin{align}
\begin{split}
\mathrm{Tr} (Q_t) &= \left( -i t^{-1}\frac{dt}{d\omega} \right) \\
&= \mathrm{Tr} \left( -i U^\dagger \frac{dU}{d\omega} +i  V^\dagger \frac{dV}{d\omega} -i \Sigma^{-1} \frac{d\Sigma}{d\omega} \right) \\
&= \sum_n \frac{d\theta_n^{(t)}}{d\omega} - i \frac{d\ln(\sigma_n^{(t)})}{d\omega} \ .
\end{split}
\label{eq:trQt}
\end{align}
Please note that the channel delay times $d\theta_n^{(t)}/d\omega$  correspond to the delay times of transmission eigenchannels and thus they differ from the real part of the $Q_t$-eigenvalues. However, the invariance of the trace under similarity transforms used in the derivation above keeps their sum the same.

Since we know that $t' = t^T$, it follows that $Q_{t'} = t^{\prime \, -1} Q_t^T t'$. Thus, $Q_{t'}$ is a similarity transform of the transpose of $Q_t$ which implies that they also share the same eigenvalues and thus their trace is the same. In case of the reflection matrix we, however, do not have a direct relation between $r$ and $r'$ and hence there is no direct relation between the eigenvalues of $Q_r$ and $Q_{r'}$, but unitarity still enforces that the sum of the corresponding channel delay times in \eqref{eq:trQ} is equal to the sum of transmission channel delay times. With the help of Eq.~\eqref{eq:trQt} and similar trace expresions for $Q_{t'}$, $Q_{r}$ and $Q_{r'}$, Eq.~\eqref{eq:trQ} can then finally be written as
\begin{align}
\begin{split}
\mathrm{Tr} (Q) &= \mathrm{Re} \left[ \mathrm{Tr} (Q_r) + \mathrm{Tr} (Q_{r'}) \right] \\
&= \mathrm{Re} \left[ \mathrm{Tr} (Q_t) + \mathrm{Tr} (Q_{t'}) \right] \\
&= 2 \mathrm{Re} \left[ \mathrm{Tr} (Q_t) \right] .
\end{split}
\label{eq:trQ_final}
\end{align}
It is important to note that there is no relation between the transmission and reflection delay times for \textit{single eigenchannels}. This can easily be seen by comparing the expressions
\begin{equation}
\frac{d\theta_n^{(t)}}{d\omega} = \frac{1}{i} \left[ 
\vec{u}_n^{\, (t,r') \dagger} \frac{d\vec{u}_n^{\, (t,r')}}{d\omega} - \vec{v}_n^{\, (t,r) \dagger} \frac{d\vec{v}_n^{\, (t,r)}}{d\omega}  \right] \neq \frac{d\theta_n^{(r)}}{d\omega}
\end{equation} 
since the left eigenvectors do not coincide, i.e., $\vec{u}_n^{\, (t)}~\neq~\vec{u}_n^{\, (r)}$. However, unitarity implies that the sum of delay times in transmission for a single eigenchannel is equal to the sum of delay times in reflection, i.e.,
\begin{equation}
\frac{d\theta_n^{(t)}}{d\omega} + \frac{d\theta_n^{(t')}}{d\omega} = \frac{d\theta_n^{(r)}}{d\omega} + \frac{d\theta_n^{(r')}}{d\omega} \ .
\end{equation}

\section{Influence of absorption in the experiment}

\subsection{Empty cavity}

\begin{figure}
\includegraphics[width=8.5cm]{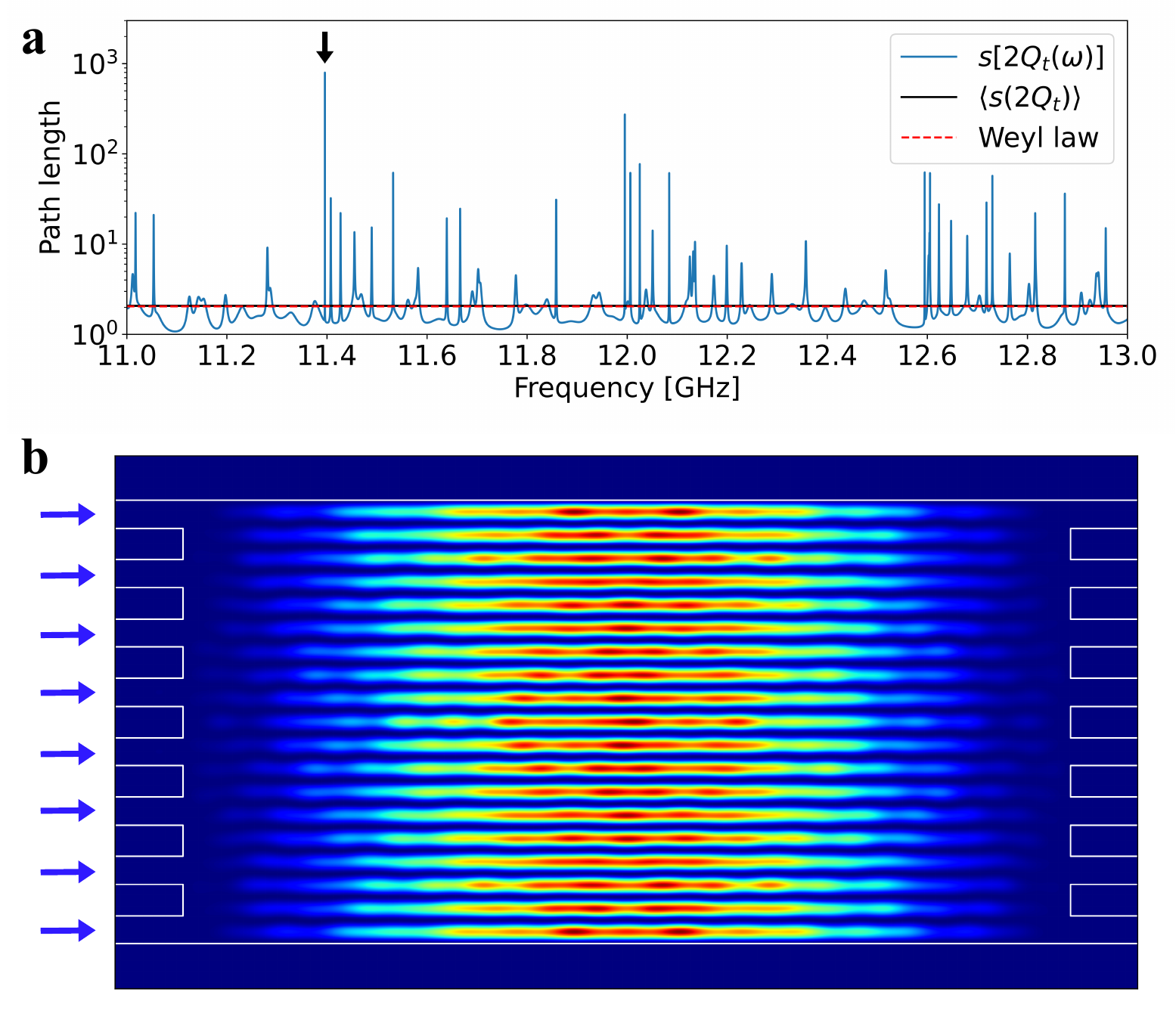}
\caption{\label{fig_distrib_eigen_delay_times}\textbf{a}, Semi-logarithmic plot of the mean path length spectrum for the empty cavity found in simulations, where the black arrow denotes the frequency at which the intensity distribution in \textbf{b} is shown. \textbf{b}, Intensity distribution of a transversally trapped transmission eigenstate at 11.42 GHz giving rise to a mean path length as large as 800 m.}
\end{figure}

We have seen in Fig.~2 of the main text that the estimated mean path length is underestimated for an empty cavity, whereas it almost reaches the theoretical prediction as soon as a few scatterers are added to our system. This underestimation can be explained by the existence of very long delay times in the empty cavity for which the impact of absorption is stronger. 
This is illustrated by the spectrum of the mean path length found in simulations in absence of absorption [see Fig. \ref{fig_distrib_eigen_delay_times}a]. Strong peaks corresponding to mean path lengths of a few hundred meters are found which are caused by bouncing orbits between the top and bottom interfaces of the cavity as shown in Fig. \ref{fig_distrib_eigen_delay_times}b. Due to their very long dwell times in the cavity, weak absorption already affects these states strongly giving rise to a dampened and statistically inhomogeneous intensity distribution which results in an underestimation of the mean path length.

\subsection{Ballistic and diffusive regime}

\begin{figure}
\includegraphics[width=8.5cm]{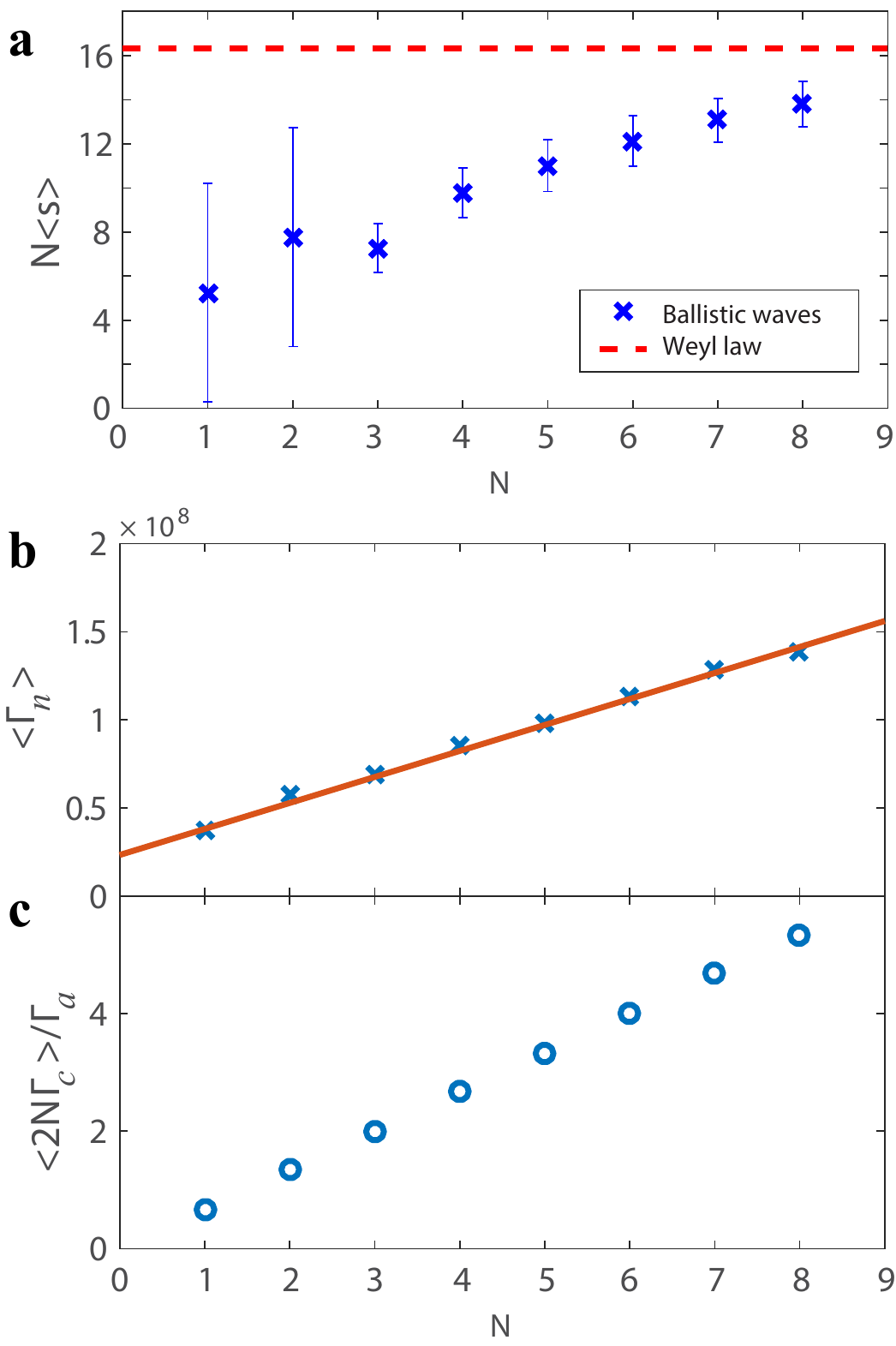}
\caption{\label{fig_variation_N_diffusive}\textbf{a}, Variation of $N\langle s \rangle$ as a function of the number of connected ports $N$ for diffusive waves with 20 scatterers inside the cavity. \textbf{b}, Average total linewidth $\langle \Gamma_n \rangle$ as a function of the number of connected ports $N$. The extrapolation for $N \rightarrow 0$ gives the linewidth associated to uniform absorption over the volume of the sample, $\Gamma_a$. \textbf{c}, Variation of the ratio $\langle \tilde{\Gamma}_n \rangle / \Gamma_a$ with $N$, where $\langle \tilde{\Gamma}_n \rangle = \langle 2 N \Gamma_c \rangle$ with $\langle \Gamma_c \rangle$ being the linewidth associated to the single port losses.}
\end{figure}

In contrast to the empty cavity, the presence of scatters fully randomizes incoming waves in the ballistic and diffusive regime which then does not allow the existence of such orbits yielding a noticeably better estimate of the mean path length in the presence of absorption. To further explain the experimentally observed enhanced robustness to absorption in the diffusive regime compared to the empty cavity, we consider a varying number of ports attached to the system, increasing from $N = 1$ to $N = 8$. For $N < 8$, the unused ports are disconnected and behave as metallic boundary conditions. To conveniently estimate the impact of absorption as a function of $N$, we consider the quantity $N\langle s\rangle = N c_0 \langle \bar{t}_\textrm{WS} \rangle$, which, in the absence of absorption, should take the same value for all systems with a different number of ports $N$.  

In the diffusive regime, as shown in Fig.~\ref{fig_variation_N_diffusive}a, the values for $N c_0 \langle \bar{t}_\textrm{WS} \rangle$ converge towards the theoretical prediction only for $N = 8$. This corresponds to the number of ports for which losses through incoming and outgoing channels dominate the dissipative losses within the sample such that $s[2Q_t(\omega)]$ provides a robust estimation of the mean path length. This can be shown by separating these two sources of losses using a decomposition of the average linewidths as a sum of losses through the ports $\langle \tilde{\Gamma}_n \rangle$ and uniform absorption $\Gamma_a$, i.e., $\langle\Gamma_n\rangle =\langle\tilde{\Gamma}_n\rangle + \Gamma_a$. The average losses through the $2N$ ports connected to the cavity scale linearly with the losses through a single channel $\langle \Gamma_c \rangle$, i.e., $\langle\tilde{\Gamma}_n\rangle = \langle 2N \Gamma_c \rangle$. To extract the values of $\Gamma_a$ and $\langle \Gamma_c \rangle$, we use that the average transmission through the sample in the time domain decreases exponentially such as $\langle |t_{ba}(t)|^2 \rangle = \exp(-\langle \Gamma_n \rangle t)$, where $\langle |t_{ba}(t)|^2 \rangle$ is found from an inverse Fourier transform of spectra of transmission coefficients $t_{ba}(\omega)$. We therefore find $\langle \Gamma_n \rangle$ for each $N$ and we fit this curve as $\langle \Gamma_n \rangle = \langle 2N \Gamma_c \rangle + \Gamma_a$ [see Fig. \ref{fig_variation_N_diffusive}b] which yields $\langle \Gamma_c \rangle \sim 7.8$ MHz and $\Gamma_a \sim 23$ MHz. For $N = 8$, $\langle 2N \Gamma_c \rangle = 125$ MHz and the ratio $\langle 2N{\Gamma}_c \rangle /\Gamma_a$ shown in Fig. \ref{fig_variation_N_diffusive}c is equal to 5.4. In this case, losses through ports indeed dominate so that the estimation of the mean path length is robust to the uniform absorption present in the experiment.

\subsection{Localized regime}

Just like in the diffusive regime, we have seen in the main text that the transmission matrix provides a reliable estimate of the Weyl prediction also in the localized regime. This is the case even though the probability distribution of the eigenchannel delay times is substantially broadened, i.e., there exist very short as well as very long paths. In the following, we provide an explanation of the robustness of the mean path length to the absorption present in the experimental setup using its connection with the density of states (DOS).

\begin{figure}
\includegraphics[width=8.5cm]{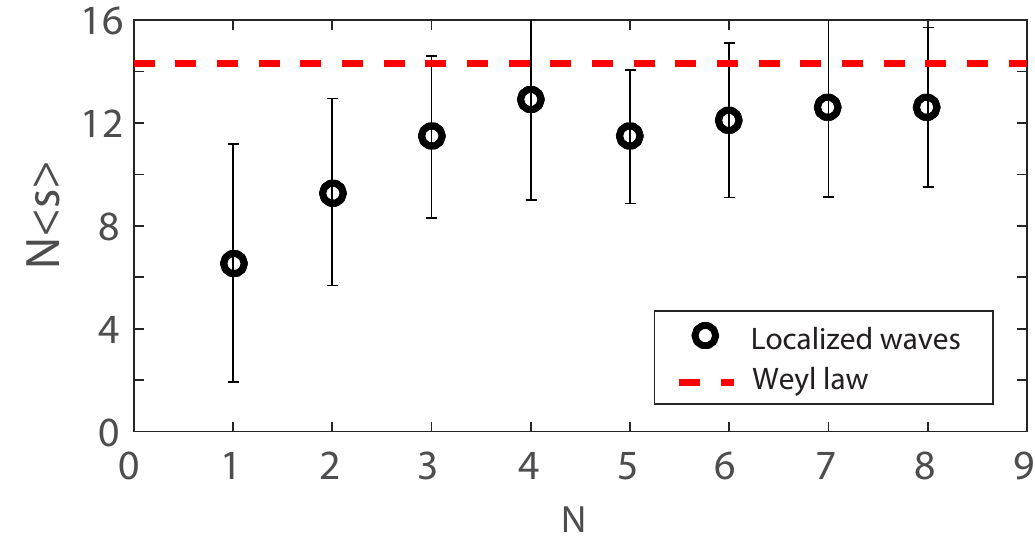}
\caption{\label{fig_variation_N_localized} Variation of $N\langle s \rangle$ as a function of the number of connected ports $N$ for localized waves with 250 scatterers inside the cavity.}
\end{figure}

 We employ the connection between the mean Wigner-Smith time delay operator and the density of states (DOS) \cite{Schwinger1951,Krein1962,Birman1992,Iannaccone1995}
\begin{equation}
\mathrm{Tr}[Q(\omega)] = 2\pi \rho(\omega)
\label{eq:trQ_DOS}
\end{equation}
to explain the robustness of Tr($Q_t$). The DOS $\rho(\omega)$ can be decomposed into a superposition of individual contributions from quasi-normal modes (QNM) of the system. These QNMs (or resonances) are associated with complex eigenfrequencies $\tilde\omega_n=\omega_n-i\Gamma_n/2$, where $\omega_n$ is the central frequency of the $n$-th mode and $\Gamma_n$ is its linewidth, such that \cite{Breit1936}
\begin{align}
\begin{split}
\rho(\omega) &= \frac{1}{\pi} \sum_n \rho_n(\omega) \\
&= \frac{1}{\pi} \sum_n \frac{\Gamma_n/2}{(\omega-\omega_n)^2+(\Gamma_n/2)^2}\,.
\end{split}
\label{eq:DOS}
\end{align}
These eigenfrequencies $\tilde\omega_n$ are also the poles of the scattering matrix. The contributions of a mode $\rho_n (\omega)$ satisfies $\int_0^\infty d\omega\rho_n (\omega) =1$ so that the average DOS is independent of absorption. Hence, adding absorption to our system results in broadening of the linewidths, $\Gamma_n \rightarrow \Gamma_n +\Gamma_a$, where $\Gamma_a$ is the linewidth associated to the losses within the sample.

The transmitted speckle pattern at the output for a spectrally isolated mode does not depend on the position of the excitation so that the transmission matrix is of unit rank \cite{Davy2012,Leseur2014,Davy2018}. Measuring the transmission through a single channel ($N = 1$) then yields the DOS using its connection to Tr($Q_t$). When several resonances overlap spectrally, however, the number of ports $N$ has to be larger than the number of resonances contributing to the DOS. We now quantify the number of ports that are required for an accurate estimation of the DOS. 

Following Breit-Wigner theory, the modal expansion of the transmission matrix is 
\begin{equation}
t(\omega)= -i W_R (\omega-\tilde\omega)^{-1} W_L^T \,
\end{equation}
where $W_R$ and $W_L$ are the projection of the eigenstates of the open system onto the right and the left channels. Furthermore, $\tilde{\omega}_n = \omega_n- i\Gamma_n/2$ are the complex eigenfrequencies of the eigenstates of the open system, i.e., the poles of the scattering matrix, where $\omega_n$ is the central frequency of the $n$-th mode and $\Gamma_n$ is the corresponding linewidth. The $W$-matrices are assumed to be independent of frequency and of dimension $N \times M$, where $N$ is the number of incoming channels and $M$ is the number of modes of the open system. Denoting their pseudo-inverses as $W_L^{-1}$ and $W_R^{-1}$ we can write 
\begin{align}
&t^{-1} = i {W_L^T}^{-1} (\omega-\tilde{\omega}) W_R^{-1} \ , \\
&\frac{dt}{d\omega} = i W_R (\omega-\tilde{\omega})^{-2} W_L^T \ ,
\end{align}
and thus
\begin{equation}
Q_t = i {W_L^T}^{-1} (\omega-\tilde\omega)^{-1} W_L^T \ ,
\label{eq:Q_t}
\end{equation}
where we have assumed that $W_R^{-1} W_R = \id$, i.e., that $W_R^{-1}$ is the left-inverse of $W_R$. The latter is only true if $M < N$ and rank$(W_R) = M$ in which case $W_R$ features orthonormal columns which results in $W_R^{-1}$ being the left inverse of $W_R$. Equation~\eqref{eq:Q_t} now shows that modes with resonances far from $\omega$ only weakly contribute to the transmission matrix. The size of $(\omega-\tilde\omega)^{-1}$ can hence be restricted to the participation ratio of modal contributions $\rho_n (\omega)$ to the DOS,
$M' = \left[ \sum_n \rho_n (\omega) \right]^2/[\Sigma_n \rho_n^2 (\omega)]$, so that  $W_L$ and $W_R$ become of dimension $N \times M'$. $M'$ is obviously linked to the modal overlap $\delta = \langle \Gamma_n \rangle / \Delta \omega$, where $\Delta \omega$ is the level spacing, $\Delta \omega=\langle \omega_{n+1}-\omega_n \rangle$. For $M' < N$, the similarity-invariance of the trace operator gives 
\begin{align}
\begin{split}
\mathrm{Re} \left[ \mathrm{Tr}(Q_t) \right] &= -\mathrm{Im} \left[ \mathrm{Tr} (\omega-\tilde{\omega})^{-1} \right] \\
&= \sum_n \frac{\Gamma_n/2}{(\omega-\omega_n )^2+(\Gamma_n/2)^2 } = \pi \rho(\omega)
\end{split}
\label{eq:trQt_DOS}
\end{align}
where we have used in the first step that $W_L^{-1}$ is the left-inverse of $W_L$ in case of $M' < N$. Inserting now Eq.~\eqref{eq:trQt_DOS} into Eq.~\eqref{eq:trQ_DOS} finally yields
\begin{equation}
\mathrm{Tr}(Q) =  2 \mathrm{Re} \left[ \mathrm{Tr}(Q_t) \right] .
\label{eq:trQ_trQt_DOS}
\end{equation}
This result now shows that the mean path length is not only well estimated from measurement of the transmission matrix only, but can also be well estimated in the presence of absorption as long as $M' < N$. However, $M'$ increases with absorption and when $M' > N$, the pseudo-inverse coupling matrices are no longer left-inverse, i.e., $W_{L,R}^{-1} W_{L,R} \neq \id$, which results in an under-estimated mean path length, i.e., $\mathrm{Re} \left[ \mathrm{Tr}(Q_t) \right] < \rho(\omega)$. Last, we want emphasize that this result is not only valid in the presence for absorption, but it also applies in the case of incomplete channel control.

In non-dissipative systems, the modal overlap $\delta$ of localized waves is small and $\langle M'\rangle \sim 1$. However, as already explained above, absorption in our experimental setup can give rise to an increase of these two quantities. Looking at our measurements through a cavity with 280 scatterers, we find that $\langle M' \rangle \approx 3.2$ so that $M' < N$ and the estimation of the mean path length is robust. The impact of the modal overlap is illustrated in Fig.~\ref{fig_variation_N_localized} in which the transmission matrix is measured with a number of ports attached to the system increasing from $N = 1$ to $N = 8$. $s_\text{theo}$ is seen to be accurately estimated for $N > 3$, in agreement with $\langle M' \rangle \approx 3.2$. In this small modal overlap regime, the first eigenchannels correspond to the contribution of single localized long-lived modes with isolated resonances. For these eigenchannels, values as large as $t_n^{(t)} = 100$ ns corresponding to travel path of 30~m in a 0.5~m long-cavity are found. Despite such long paths, we observe in Fig.~\ref{fig_variation_N_localized} that the mean path length estimated with respect to the number of connected ports, $N c_0 \langle \bar{t}_\textrm{WS} \rangle$, reaches a plateau already when the number of connected ports to the cavity satisfies $N > 3$. 

It is worth pointing out that the transmission eigenvalues $\tau_n$ do not contribute to the real part of the eigenvalues of $Q_t$ since the latter only contains the frequency-derivatives of the singular vectors of $t$ which allows an accurate estimation of the mean path length even in such localizing systems featuring very low transmission.

\section{Convergence towards the most direct path in the limit of strong absorption}

\begin{figure}
\includegraphics[width=8.5cm]{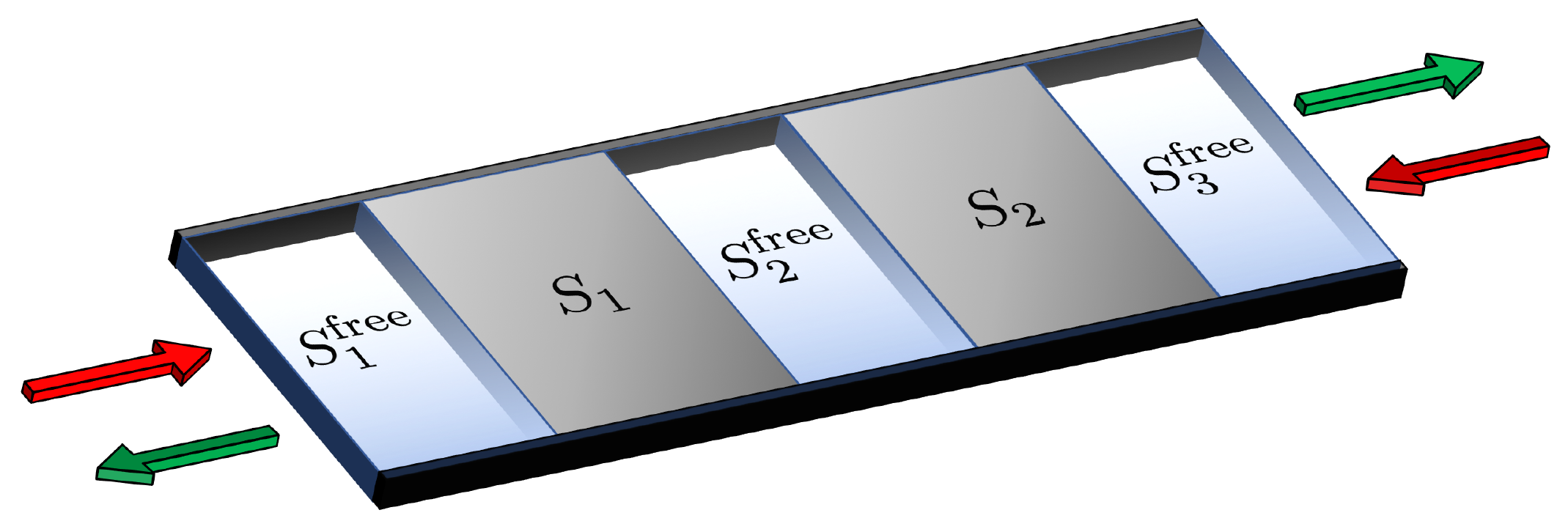}
\caption{Sketch of the two-dimensional toy-model, where $S_j^\mathrm{free}$ with $j=1,2,3$ are the scattering matrices of the free space sections and $S_1$ and $S_2$ are scattering matrices of the two considered scattering regions.}
\label{fig_toymodel}
\end{figure}

To investigate the effect of strong absorption on all operators involved in Eq.~\eqref{eq:trQ_final}, we consider the two-dimensional toy-model shown in Fig.~\ref{fig_toymodel} which consists of two scattering regions with scattering matrices
\begin{equation}
S_1 = \begin{pmatrix}
r_1 & t_1' \\
t_1 & r'_1
\end{pmatrix} , \quad 
S_2 = \begin{pmatrix}
r_2 & t_2' \\
t_2 & r'_2
\end{pmatrix}
\end{equation}
separated by free space described by the scattering matrix $S_2^\mathrm{free}$, where we also add a free space section described by $S_1^\mathrm{free}$ ($S_3^\mathrm{free}$) in front of the first (after the second) scattering region for generality. The corresponding scattering matrices for these free space sections are given by
\begin{equation}
S_j^\mathrm{free} = \begin{pmatrix}
0 & P_j \\
P_j & 0
\end{pmatrix} ,
\end{equation}
where the propagation matrices are defined as
\begin{equation}
P_j = \mathrm{diag} \left( e^{i k_{x,n} L_j} \right) \ .
\end{equation}
Here, $k_{x,n} = \sqrt{k^2-k_{y,n}^2}$ are the longitudinal propagation constants of the waveguide modes, $k = 2 \pi/\lambda$ is the free space wave vector, $k_{y,n} = n \pi / W$ are the transverse wave vectors of the modes in a waveguide of width $W$ and $L_j$ are the lengths of the corresponding free space sections.

\subsection{Trace of $Q_t$}

To calculate the transmission matrix of such a system, we employ the Feynman path integral formulation and sum over all possible paths which yields
\begin{align}
\begin{split}
t =\, &P_1 t_1 P_2 t_2 P_3 \\
&+ P_1 t_1 P_2 r_2 P_2 r'_1 P_2 t_2 P_3 \\
&+ P_1 t_1 P_2 r_2 P_2 r'_1 P_2 r_2 P_2 r'_1 P_2 t_2 P_3 +\ \ldots \\
=\, & P_1 t_1 P_2 \sum_{n=0}^\infty (r_2 P_2 r'_1 P_2)^n t_2 P_3 \\
=\, & P_1 t_1 P_2 (\id - r_2 P_2 r'_1 P_2)^{-1} t_2 P_3 \ ,
\end{split}
\end{align}
Hence
\begin{equation}
t^{-1} = P_3^{-1} t_2^{-1} (\id - r_2 P_2 r'_1 P_2) P_2^{-1} t_1^{-1} P_1^{-1}
\end{equation}
and using these expressions to calculate the trace of $Q_t = -i t^{-1} dt/dk$ then gives
\begin{align}
\mathrm{Tr}(Q_t) = \mathrm{Tr}(Q_t)_\mathrm{direct} + \mathrm{Tr}(Q_t)_\mathrm{multi}
\end{align}
with
\begin{align}
\mathrm{Tr}(Q_t)_\mathrm{direct} = \mathrm{Tr} \bigg[ 
&-i P_1^{-1} \frac{dP_1}{dk} \label{eq:trQt_direct}
-i t_1^{-1} \frac{dt_1}{dk} \\
&-i P_2^{-1} \frac{dP_2}{dk}
-i t_2^{-1} \frac{dt_2}{dk} 
-i P_3^{-1} \frac{dP_3}{dk} \bigg] , \nonumber
\end{align}
being the time-delay caused by the direct path and
\begin{align}
\begin{split}
\mathrm{Tr}(Q_t)_\mathrm{multi} = \mathrm{Tr} \bigg[ -i (\id &- r_2 P_2 r'_1 P_2) \\
&\times \frac{d (\id - r_2 P_2 r'_1 P_2)^{-1}}{dk} \bigg]
\end{split}
\label{eq:trQt_multi}
\end{align}
corresponds to the contributions from the multiply scattered paths, where we have used the linearity of the trace and its invariance under cyclic permutations of matrices. 

We now add uniform absorption by adding an imaginary part to the total wave vector, i.e., $k \to k+i\kappa$, where $\kappa = n_i k$ with $n_i$ being the uniform imaginary part of the refractive index. This imaginary part then causes a change of the propagation constants 
\begin{equation}
k_{x,n} \to \tilde{k}_{x,n} \left(1+i n_i k^2 / \tilde{k}_{x,n}^2 \right)
\end{equation}
 with $\tilde{k}_{x,n} = \sqrt{(1-n_i^2) k^2 - k_{y,n}^2}$ in case of $n_i \ll 1$. Thus, we get
 \begin{equation}
 P_j \to \mathrm{diag} \left( e^{i \tilde{k}_{x,n} L_j} e^{-n_i L_j k^2/\tilde{k}_{x,n}}\right) \ ,
  \label{eq:pj_absorption}
 \end{equation}
 where its derivative evaluates to
 \begin{equation}
 \frac{d P_j}{dk} \to \mathrm{diag} \left( \frac{i k L_j}{\tilde{k}_{x,n}} - n_i L_j \left[ \frac{2k}{\tilde{k}_{x,n}} - \frac{k^3}{\tilde{k}_{x,n}^3} \right] \right) P_j \ .
   \label{eq:pj_deriv_absorption}
 \end{equation}
 Taylor-expanding now the inverse matrix expression in \eqref{eq:trQt_multi} yields
 \begin{align}
 \begin{split}
 (\id - r_2 P_2 r'_1 P_2)^{-1} = \id &+ r_2 P_2 r'_1 P_2 \\
 &+ (r_2 P_2 r'_1 P_2)^2 +\ \ldots
 \end{split}
 \end{align}
 and due to \eqref{eq:pj_deriv_absorption} we find that
 \begin{equation}
 \frac{d(\id - r_2 P_2 r'_1 P_2)^{-1}}{dk} \propto \mathcal{O}(P_2^2) \ .
 \end{equation}
 This reveals now that $\mathrm{Tr}(Q_t)_\mathrm{multi} \propto \mathcal{O}(P_2^2) \to 0$ in the limit of strong absorption due to the exponential decay in \eqref{eq:pj_absorption} and thus
 \begin{equation}
 \mathrm{Tr}(Q_t) \to \mathrm{Tr}(Q_t)_\mathrm{direct} \ .
 \label{eq:trQt_absorption}
 \end{equation}
 This means that the transmission delay times converge to those of the most direct paths in the limit of strong absorption which is already reached for $n_i \ll 1$ due to the fast exponential decrease in \eqref{eq:pj_absorption}. In a less complex one-dimensional system featuring barriers of finite height, it can be even shown that $Q_t$ reduces to the optical path length of the direct path through this system if absorption is strong. It is also worth pointing out that $\mathrm{Tr}(Q_t)_\mathrm{direct}$ is independent of absorption in the sense that the exponential decay drops out in \eqref{eq:trQt_direct}, but absorption still enters in form of a correction term in \eqref{eq:pj_deriv_absorption}. However, since usually $n_i \ll 1$, these corrections are small and one can approximate $\tilde{k}_{x,n} \approx k_{x,n}$ and $dP_j/dk \approx \mathrm{diag}(i k L_j/k_{x,n})$ where the latter expression yields the free space propagation delay time in the absence of absorption in \eqref{eq:trQt_direct}.

\subsection{Trace of $Q_r$}

Next, we investigate the effect of absorption on $Q_r$. To set up the reflection matrix of our toy-model system, we again sum over all possible paths and get
 \begin{align}
 \begin{split}
r =\, &P_1 r_1 P_1 \\
&+ P_1 t_1 P_2 r_2 P_2 t'_1 P_1  \\
&+ P_1 t_1 P_2 r_2 P_2 r'_1 P_2 r_2 P_2 t'_1 P_1 +\ \ldots \\
=\, &P_1 r_1 P_1 + P_1 t_1 P_2 r_2 P_2 \sum_{n=0}^\infty (r'_1 P_2 r_2 P_2)^n t'_1 P_1 \\
=\, &P_1 r_1 P_1 + P_1 t_1 P_2 r_2 P_2 (\id - r'_1 P_2 r_2 P_2)^{-1} t'_1 P_1
 \end{split}
\end{align}
To calculate the inverse of $r$ we first rewrite the above expression as follows
\begin{align}
\begin{split}
r =\, P_1 r_1 P_1 \big[ \id + &P_1^{-1} r_1^{-1} t_1 P_2 r_2 P_2 \\
&\times (\id - r'_1 P_2 r_2 P_2)^{-1} t'_1 P_1 \big] . 
\end{split}
\end{align}
Next, we assume strong absorption which causes the term in the square brackets to be small thus enabling us to approximate its inverse by $(1+x)^{-1} \approx 1-x$ which yields
\begin{align}
\begin{split}
r^{-1} \approx \big[ \id - &P_1^{-1} r_1^{-1} t_1 P_2 r_2 P_2 \\
&\times(\id - r'_1 P_2 r_2 P_2)^{-1} t'_1 P_1 \big] P_1^{-1} r_1^{-1} P_1^{-1} \ .
\end{split}
\end{align}
Following the same strategy as above and calculating $dr/dk$ and then $Q_r$, we find that 
\begin{align}
\mathrm{Tr}(Q_r) = \mathrm{Tr}(Q_r)_\mathrm{direct} + \mathrm{Tr}(Q_r)_\mathrm{multi} \ ,
\end{align}
with the contributions from the direct path being
\begin{align}
\begin{split}
\mathrm{Tr}(Q_r)_\mathrm{direct} = \mathrm{Tr} \bigg[ 
&-i P_1^{-1} \frac{dP_1}{dk} \\
&-i r_1^{-1} \frac{dr_1}{dk}
-i P_1^{-1} \frac{dP_1}{dk} \bigg] ,
\end{split}
\label{eq:trQr_direct}
\end{align}
where the contributions from the mutliply scattered paths are again
\begin{align}
\mathrm{Tr}(Q_r)_\mathrm{multi} = \mathcal{O}(P_2^2) \ .
\label{eq:trQr_multi}
\end{align}
In the limit of strong absorption, we thus again find that $\mathrm{Tr}(Q_r)_\mathrm{multi} \to 0$ which yields
\begin{equation}
\mathrm{Tr}(Q_r) \to \mathrm{Tr}(Q_r)_\mathrm{direct} \ .
\label{eq:trQr_absorption}
\end{equation}
Hence, for strong absorption and $n_i \ll 1$, the eigenchannel delay times in reflection reduce to the delay times of the most direct paths in the absence of absorption as was the case for the time delay operator containing the transmission matrix.

\begin{figure*}
\centering
\includegraphics[width=18cm]{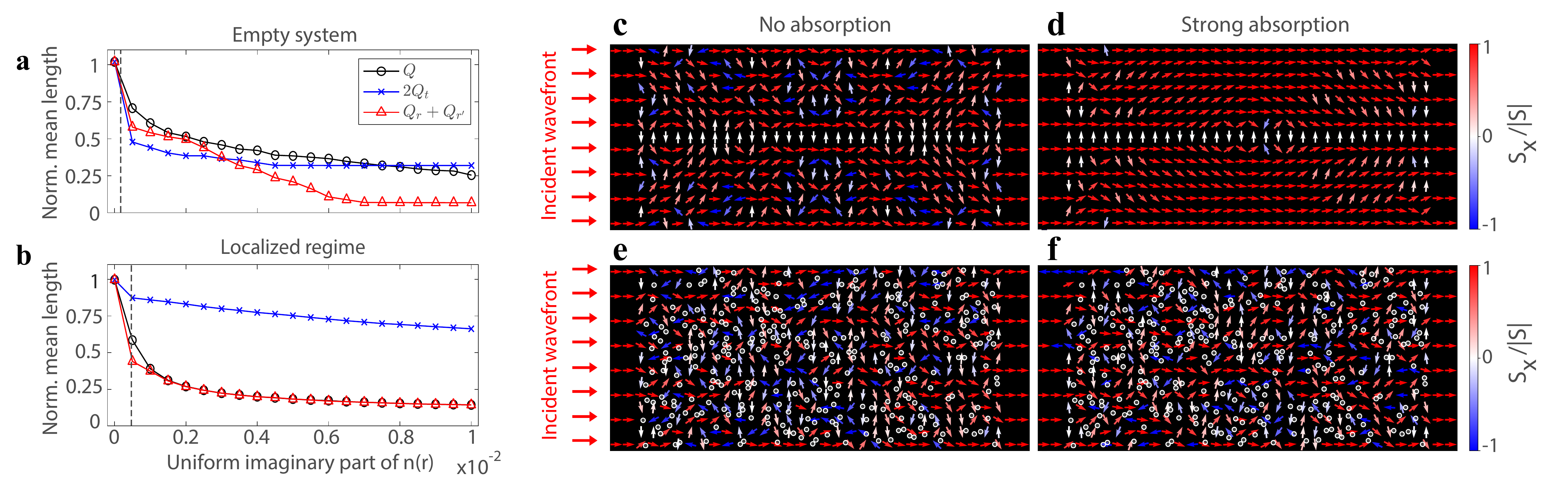}
\caption{ \textbf{a},\textbf{b}, Normalized mean path lengths obtained in numerical simulations by $\langle s(\mathcal{O}) \rangle / \langle s_\mathrm{theo} \rangle$ with $\mathcal{O}$ representing the corresponding operators in the legend. The ratios of these quantities with the dissipationless value $\langle s_{\text{theo}} \rangle$ obtained from Eq. (2) monotonically decrease with increasing absorption strength both for the empty cavity \textbf{a} ($\langle s_\text{theo} \rangle = 2.05$ m) and in the localized regime  \textbf{b} ($n_\text{scat} = 280$, $\langle s_\text{theo} \rangle = 1.76$ m). The vertical dashed lines indicate the absorption strength for which the frequency-averaged conductance in the simulation is equal to the experimentally measured one.
\textbf{c-f}, Spatial distribution of the Poynting vector for the highest transmitting eigenchannel of the empty cavity (\textbf{c},\textbf{e}) and a localized sample (\textbf{c},\textbf{e}) in the the absence of absorption (\textbf{c},\textbf{d}) and in case of strong absorption (\textbf{e}, \textbf{f}). The colors correspond to the longitudinal component of the Poynting vector normalized by its magnitude.}
\label{fig:SM_direct_path}
\end{figure*}

\subsection{Trace of $Q$}

Last, we investigate the effect of absorption on Q which contains the full scattering matrix of the system
\begin{equation}
S = \begin{pmatrix}
r & t' \\
t & r'
\end{pmatrix} .
\end{equation}
The scattering matrix $S$ can then be inverted blockwise
\begin{equation}
S^{-1} = \begin{pmatrix}
r^{-1} + r^{-1} t' M t r^{-1} & -r^{-1} t' M \\
- M t r^{-1} & M
\end{pmatrix}
\end{equation}
with $M = (r' - t r^{-1} t')^{-1}$ which is valid if $r$ and $M$ are non-singular. The trace of $Q = -i S^{-1} dS/dk$ is then given by
\begin{align}
\mathrm{Tr} (Q) = \mathrm{Tr} \Bigg[ &-i r^{-1} \frac{dr}{dk} -i r^{-1} t' M t r^{-1} \frac{dr}{dk} \\
&+i r^{-1} t' M \frac{dt}{dk} + i M t r^{-1} \frac{dt'}{dk} -i M \frac{dr'}{dk} \Bigg] . \nonumber
\end{align}
Since absorption causes an exponential decrease in transmission, strong absorption will lead to $t \to 0$ and $t' \to 0$. In this limit, $M \to r^{\prime -1}$ and 
\begin{equation}
\mathrm{Tr}(Q) \to \mathrm{Tr}(Q_r) + \mathrm{Tr}(Q_{r'}) \ .
\label{eq:trQ_absorption}
\end{equation}

\subsection{Numerical results}

To investigate the influence of strong absorption in more detail, we perform numerical simulations of the experimental setup between 11 and 13~GHz and introduce uniform absorption of increasing strength by adding a progressively increasing imaginary part to the effective refractive index of the cavity. The results in the empty cavity and localized regime are shown in Fig.~\ref{fig:SM_direct_path}a,b, where each data point is the result of a frequency- and configuration average over 200 random configurations. In both cases, we observe that the path lengths obtained with the different operators decrease with respect to absorption.

In the empty system, the transmission matrix and reflection matrices as well as the full scattering matrix provide mean path lengths that are much smaller than Weyl prediction in the strong absorption limit. To understand the rapid decrease of the path lengths for an increasing absorption strength, we show in Fig.~\ref{fig:SM_direct_path} the spatial Poynting vector distribution of the eigenstate featuring the highest transmission in the empty cavity at 10 GHz. In the absence of absorption, complex patterns in the Poynting vector distribution are visible (see Fig.~\ref{fig:SM_direct_path}c). However adding strong absorption causes the flux to align with the positive x-direction (red color) which illustrates the convergence towards the most direct path from the input to the output ports (see Fig.~\ref{fig:SM_direct_path}e). 

To estimate the most direct path in the empty cavity, we use that the time-delay of a straight section of length $\ell$ without any scatterers is $\ell k/k_{x,n}$ with $k = 2\pi/\lambda$ being the total free space wavevectora and $k_{x,n} = \sqrt{k^2-(n\pi/W)^2}$ being the propagation constant of the $n$-th mode in a section of width $W$. In the limit of strong absorption, only the most direct path will survive which is given by the time-delay of the lowest transverse waveguide mode in the corresponding sections and thus our estimate reads 
\begin{equation}
\langle L_\mathrm{direct}^\mathrm{empty} (k) \rangle \approx \left\langle \frac{2L_\mathrm{lead} k}{k_{x,1}^\mathrm{lead}} + \frac{L_\mathrm{scat} k}{k_{x,1}^\mathrm{scat}} \right\rangle = 0.63 \ \mathrm{m}\ ,
\end{equation}
where $L_\mathrm{lead} = 38$ mm, $k_{x,1}^\mathrm{lead} = \sqrt{k^2-(\pi/W_\mathrm{lead})^2}$, $W_\mathrm{lead} = 15.79$ mm, $L_\mathrm{scat} = 500$ mm, $k_{x,1}^\mathrm{scat} = \sqrt{k^2-(\pi/W_\mathrm{scat})^2}$, $W_\mathrm{scat} = 250$ mm and the average was performed over the frequency range of 11-13 GHz. Normalizing it now with the corresponding average mean path length  in this frequency interval $\langle s_\mathrm{theo} \rangle = 2.05$ m then yields
\begin{equation}
\frac{\langle L_\mathrm{direct}^\mathrm{empty} \rangle}{\langle s_\mathrm{theo} \rangle} \approx 0.31 \ .
\end{equation}
This value is in very good agreement with the value of $\langle s(2Q_t) \rangle / \langle s_\mathrm{theo} \rangle$ for the largest absorption strength which is slightly larger (0.32) due to the fact the direct paths can also contain contributions featuring a slight angle to the longitudinal axis causing the waves to exit the system through one of the output ports not exactly opposite to their input port. Moreover, this absorption strength might still be not strong enough, i.e., we might not have fully converged to the direct path yet.

In the localized sample, the transmission delay times in the limit of strong absorption also reduce to the ones of the most direct paths which still have to traverse the system. However, adding impenetrable metallic scatterers to the scattering region increases the length of the most direct path as can be seen by comparing the mean path length for strong absorption in Fig.~\ref{fig:SM_direct_path}b with Fig.~\ref{fig:SM_direct_path}a. The distribution of the Poynting vector in Fig.~\ref{fig:SM_direct_path}f now shows a complex pattern made of tortuous paths within the sample even in the strong absorption regime. The transmission eigenchannel delay times hence converge to a value much larger than the one associated to $\langle L_\mathrm{direct}^\mathrm{empty} \rangle$. This elongation of the direct path leads thus to an increased mean path length compared to the empty system in the strong absorption limit and thus to a better estimate of the value predicted by the Weyl law.

In contrast, the reflection delay times decrease and converge in the localizing configurations to small values which resemble the direct paths in reflection, i.e., direct reflections at the first scattering layers -- as predicted by Eq.~\eqref{eq:trQr_absorption} -- where the trace of $Q$ converges to the sum of reflection delay times due to the almost vanishing transmission [see Eq.~\eqref{eq:trQ_absorption}]. Thus, estimating the mean path length using the transmission matrix provides a much more reliable estimator of the Weyl prediction in the localized regime than its estimation based on the full scattering matrix as illustrated in Fig.~\ref{fig:SM_direct_path}b.

\begin{figure}
\centering
\includegraphics[width=8.5cm]{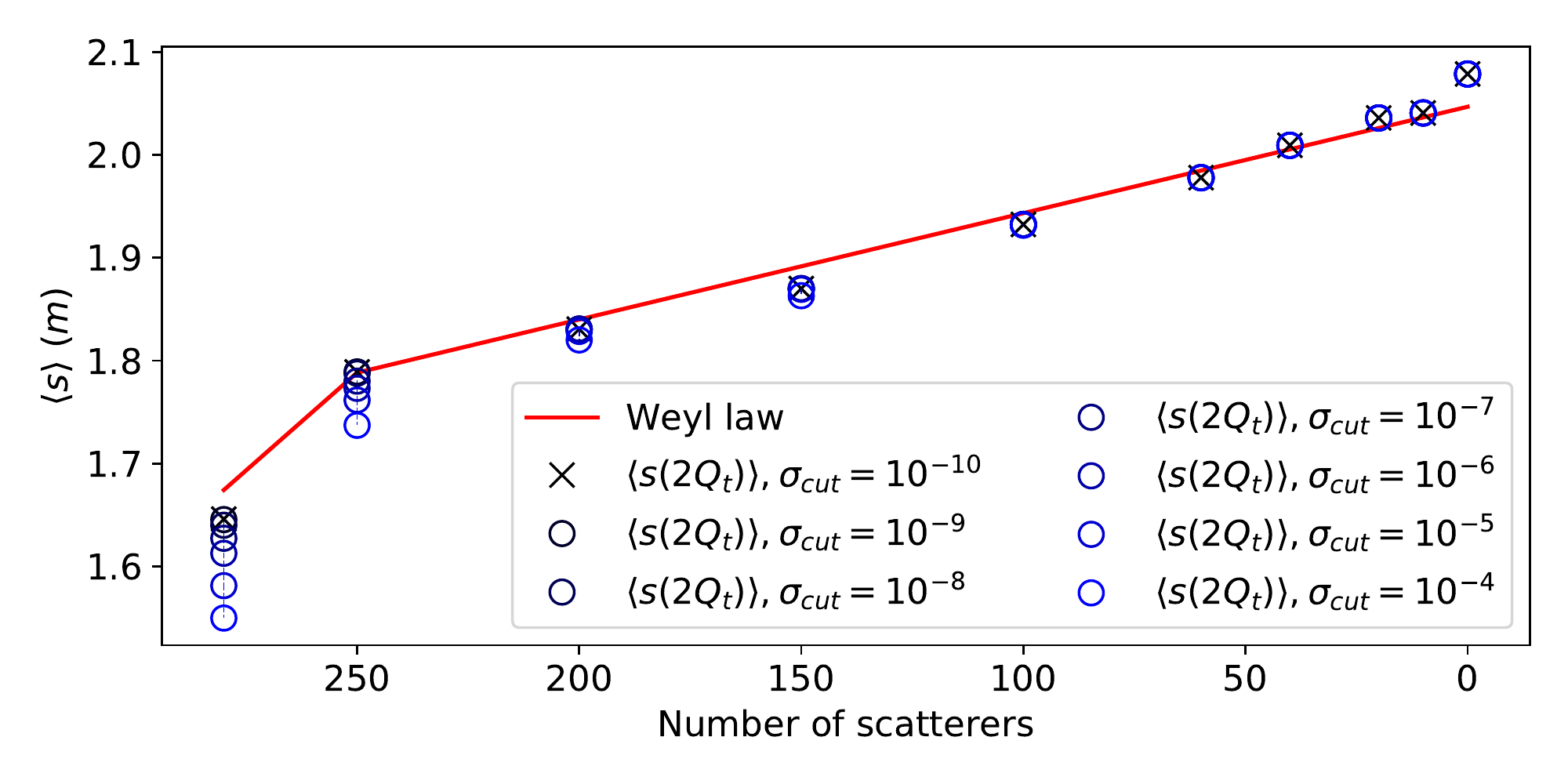}
\caption{ Mean path length $\langle s(2Q_t) \rangle$ for different singular value cut-offs $\sigma_\mathrm{cut}$ (black crosses and blue circles), where the black crosses correspond to the ones in Fig.~2a of the main text for which a cut-off of $\sigma_\mathrm{cut} = 10^{-10}$ was chosen. The red line shows the theoretical value $\langle s_\mathrm{theo} \rangle$ obtained using the Weyl law (frequency average of Eq.~2 in the main text). The deviations observed here from the theoretical values are typically decreasing, when we average not only over frequency (as done here), but also over different disorder configurations (not shown). To make sure that the system with 280 scatterers is clearly in the localized regime, the frequency range was here reduced to 11-12 GHz, as compared to the range 11-13 GHz used for all the other scatterer numbers.}
\label{fig:SM_svdcutoff}
\end{figure}

{
\section{Influence of the truncation of singular values of the TM}
In this section, we take a closer look at the influence of the cut-off applied to the singular values for computing the pseudo-inverse of the scattering, transmission and reflection matrices. Figure \ref{fig:SM_svdcutoff} shows the mean path length $\langle s (2Q_t) \rangle$ obtained from the numerical simulations for different cut-off values $\sigma_\mathrm{cut}$, where $\sigma_\mathrm{cut} = 10^{-10}$ corresponds to the value chosen in the main text. In the empty system and in the diffusive regime, the transmission is high enough such that the cut-off does not affect the mean path length. However, in the localized regime we clearly see a decrease of the mean path length with increasing cut-off, which is caused by low transmission values $\sigma < \sigma_\mathrm{cut}$ that are cut away in our SVD projection procedure.
}

\bibliographystyle{apsrev4-1}

\begin{thebibliography}{42}%
\makeatletter
\providecommand \@ifxundefined [1]{%
 \@ifx{#1\undefined}
}%
\providecommand \@ifnum [1]{%
 \ifnum #1\expandafter \@firstoftwo
 \else \expandafter \@secondoftwo
 \fi
}%
\providecommand \@ifx [1]{%
 \ifx #1\expandafter \@firstoftwo
 \else \expandafter \@secondoftwo
 \fi
}%
\providecommand \natexlab [1]{#1}%
\providecommand \enquote  [1]{``#1''}%
\providecommand \bibnamefont  [1]{#1}%
\providecommand \bibfnamefont [1]{#1}%
\providecommand \citenamefont [1]{#1}%
\providecommand \href@noop [0]{\@secondoftwo}%
\providecommand \href [0]{\begingroup \@sanitize@url \@href}%
\providecommand \@href[1]{\@@startlink{#1}\@@href}%
\providecommand \@@href[1]{\endgroup#1\@@endlink}%
\providecommand \@sanitize@url [0]{\catcode `\\12\catcode `\$12\catcode
  `\&12\catcode `\#12\catcode `\^12\catcode `\_12\catcode `\%12\relax}%
\providecommand \@@startlink[1]{}%
\providecommand \@@endlink[0]{}%
\providecommand \url  [0]{\begingroup\@sanitize@url \@url }%
\providecommand \@url [1]{\endgroup\@href {#1}{\urlprefix }}%
\providecommand \urlprefix  [0]{URL }%
\providecommand \Eprint [0]{\href }%
\providecommand \doibase [0]{http://dx.doi.org/}%
\providecommand \selectlanguage [0]{\@gobble}%
\providecommand \bibinfo  [0]{\@secondoftwo}%
\providecommand \bibfield  [0]{\@secondoftwo}%
\providecommand \translation [1]{[#1]}%
\providecommand \BibitemOpen [0]{}%
\providecommand \bibitemStop [0]{}%
\providecommand \bibitemNoStop [0]{.\EOS\space}%
\providecommand \EOS [0]{\spacefactor3000\relax}%
\providecommand \BibitemShut  [1]{\csname bibitem#1\endcsname}%
\let\auto@bib@innerbib\@empty

\bibitem [{\citenamefont {Dirac}(1943)}]{Dirac1943}%
  \BibitemOpen
  \bibfield  {author} {\bibinfo {author} {\bibfnamefont {P.}~\bibnamefont
  {Dirac}},\ }\href@noop {} {\bibfield  {journal} {\bibinfo  {journal}
  {Declassified British Report MS-D-5, Part I}\ } (\bibinfo {year}
  {1943})}\BibitemShut {NoStop}%
\bibitem [{\citenamefont {Case}\ and\ \citenamefont
  {Zweifel}(1967)}]{Case1967}%
  \BibitemOpen
  \bibfield  {author} {\bibinfo {author} {\bibfnamefont {K.~M.}\ \bibnamefont
  {Case}}\ and\ \bibinfo {author} {\bibfnamefont {P.~F.}\ \bibnamefont
  {Zweifel}},\ }\href@noop {} {\emph {\bibinfo {title} {Linear transport
  theory}}}\ (\bibinfo  {publisher} {Addison-Wesley},\ \bibinfo {year}
  {1967})\BibitemShut {NoStop}%
\bibitem [{\citenamefont {Blanco}\ and\ \citenamefont
  {Fournier}(2003)}]{Blanco2003}%
  \BibitemOpen
  \bibfield  {author} {\bibinfo {author} {\bibfnamefont {S.}~\bibnamefont
  {Blanco}}\ and\ \bibinfo {author} {\bibfnamefont {R.}~\bibnamefont
  {Fournier}},\ }\href@noop {} {\bibfield  {journal} {\bibinfo  {journal}
  {Europhys. Lett.}\ }\textbf {\bibinfo {volume} {61}},\ \bibinfo {pages} {168}
  (\bibinfo {year} {2003})}\BibitemShut {NoStop}%
\bibitem [{\citenamefont {Vasiliev}\ \emph {et~al.}(2019)\citenamefont
  {Vasiliev}, \citenamefont {Nur-E-Alam},\ and\ \citenamefont
  {Alameh}}]{Vasiliev2019}%
  \BibitemOpen
  \bibfield  {author} {\bibinfo {author} {\bibfnamefont {M.}~\bibnamefont
  {Vasiliev}}, \bibinfo {author} {\bibfnamefont {M.}~\bibnamefont
  {Nur-E-Alam}}, \ and\ \bibinfo {author} {\bibfnamefont {K.}~\bibnamefont
  {Alameh}},\ }\href@noop {} {\bibfield  {journal} {\bibinfo  {journal}
  {Energies}\ }\textbf {\bibinfo {volume} {12}},\ \bibinfo {pages} {1080}
  (\bibinfo {year} {2019})}\BibitemShut {NoStop}%
\bibitem [{\citenamefont {Frangipane}\ \emph {et~al.}(2019)\citenamefont
  {Frangipane}, \citenamefont {Vizsnyiczai}, \citenamefont {Maggi},
  \citenamefont {Savo}, \citenamefont {Sciortino}, \citenamefont {Gigan},\ and\
  \citenamefont {Di~Leonardo}}]{Frangipane2019}%
  \BibitemOpen
  \bibfield  {author} {\bibinfo {author} {\bibfnamefont {G.}~\bibnamefont
  {Frangipane}}, \bibinfo {author} {\bibfnamefont {G.}~\bibnamefont
  {Vizsnyiczai}}, \bibinfo {author} {\bibfnamefont {C.}~\bibnamefont {Maggi}},
  \bibinfo {author} {\bibfnamefont {R.}~\bibnamefont {Savo}}, \bibinfo {author}
  {\bibfnamefont {A.}~\bibnamefont {Sciortino}}, \bibinfo {author}
  {\bibfnamefont {S.}~\bibnamefont {Gigan}}, \ and\ \bibinfo {author}
  {\bibfnamefont {R.}~\bibnamefont {Di~Leonardo}},\ }\href {\doibase
  10.1038/s41467-019-10455-y} {\bibfield  {journal} {\bibinfo  {journal} {Nat.
  Comm.}\ }\textbf {\bibinfo {volume} {10}},\ \bibinfo {pages} {2442} (\bibinfo
  {year} {2019})}\BibitemShut {NoStop}%
\bibitem [{\citenamefont {Pierrat}\ \emph {et~al.}(2014)\citenamefont
  {Pierrat}, \citenamefont {Ambichl}, \citenamefont {Gigan}, \citenamefont
  {Haber}, \citenamefont {Carminati},\ and\ \citenamefont
  {Rotter}}]{Pierrat2014}%
  \BibitemOpen
  \bibfield  {author} {\bibinfo {author} {\bibfnamefont {R.}~\bibnamefont
  {Pierrat}}, \bibinfo {author} {\bibfnamefont {P.}~\bibnamefont {Ambichl}},
  \bibinfo {author} {\bibfnamefont {S.}~\bibnamefont {Gigan}}, \bibinfo
  {author} {\bibfnamefont {A.}~\bibnamefont {Haber}}, \bibinfo {author}
  {\bibfnamefont {R.}~\bibnamefont {Carminati}}, \ and\ \bibinfo {author}
  {\bibfnamefont {S.}~\bibnamefont {Rotter}},\ }\href {\doibase
  10.1073/pnas.1417725111} {\bibfield  {journal} {\bibinfo  {journal}
  {Proceedings of the National Academy of Sciences}\ }\textbf {\bibinfo
  {volume} {111}},\ \bibinfo {pages} {17765} (\bibinfo {year}
  {2014})}\BibitemShut {NoStop}%
\bibitem [{\citenamefont {Schwinger}(1951)}]{Schwinger1951}%
  \BibitemOpen
  \bibfield  {author} {\bibinfo {author} {\bibfnamefont {J.}~\bibnamefont
  {Schwinger}},\ }\href@noop {} {\bibfield  {journal} {\bibinfo  {journal}
  {Phys. Rev.}\ }\textbf {\bibinfo {volume} {82}},\ \bibinfo {pages} {664}
  (\bibinfo {year} {1951})}\BibitemShut {NoStop}%
\bibitem [{\citenamefont {Krein}(1962)}]{Krein1962}%
  \BibitemOpen
  \bibfield  {author} {\bibinfo {author} {\bibfnamefont {M.~G.}\ \bibnamefont
  {Krein}},\ }\href@noop {} {\bibfield  {journal} {\bibinfo  {journal} {Dokl.
  Akad. Nauk SSSR}\ }\textbf {\bibinfo {volume} {144}},\ \bibinfo {pages} {475}
  (\bibinfo {year} {1962})}\BibitemShut {NoStop}%
\bibitem [{\citenamefont {Birman}\ and\ \citenamefont
  {Yafaev}(1992)}]{Birman1992}%
  \BibitemOpen
  \bibfield  {author} {\bibinfo {author} {\bibfnamefont {M.~S.}\ \bibnamefont
  {Birman}}\ and\ \bibinfo {author} {\bibfnamefont {D.~R.}\ \bibnamefont
  {Yafaev}},\ }\href@noop {} {\bibfield  {journal} {\bibinfo  {journal}
  {Algebra i Analiz}\ }\textbf {\bibinfo {volume} {4}},\ \bibinfo {pages} {1}
  (\bibinfo {year} {1992})}\BibitemShut {NoStop}%
\bibitem [{\citenamefont {Iannaccone}(1995)}]{Iannaccone1995}%
  \BibitemOpen
  \bibfield  {author} {\bibinfo {author} {\bibfnamefont {G.}~\bibnamefont
  {Iannaccone}},\ }\href@noop {} {\bibfield  {journal} {\bibinfo  {journal}
  {Phys. Rev. B}\ }\textbf {\bibinfo {volume} {51}},\ \bibinfo {pages} {4727}
  (\bibinfo {year} {1995})}\BibitemShut {NoStop}%
\bibitem [{\citenamefont {Weyl}(1911)}]{Weyl1911}%
  \BibitemOpen
  \bibfield  {author} {\bibinfo {author} {\bibfnamefont {H.}~\bibnamefont
  {Weyl}},\ }\href {http://eudml.org/doc/58792} {\bibfield  {journal} {\bibinfo
   {journal} {{Nachrichten von der Gesellschaft der Wissenschaften zu
  G\"ottingen, Mathematisch-Physikalische Klasse}}\ }\textbf {\bibinfo {volume}
  {1911}},\ \bibinfo {pages} {110} (\bibinfo {year} {1911})}\BibitemShut
  {NoStop}%
\bibitem [{\citenamefont {Arendt}\ \emph {et~al.}(2009)\citenamefont {Arendt},
  \citenamefont {Nittka}, \citenamefont {Peter}, \citenamefont {Steiner},\ and\
  \citenamefont {Schleich}}]{Arendt}%
  \BibitemOpen
  \bibfield  {author} {\bibinfo {author} {\bibfnamefont {W.}~\bibnamefont
  {Arendt}}, \bibinfo {author} {\bibfnamefont {R.}~\bibnamefont {Nittka}},
  \bibinfo {author} {\bibfnamefont {W.}~\bibnamefont {Peter}}, \bibinfo
  {author} {\bibfnamefont {F.}~\bibnamefont {Steiner}}, \ and\ \bibinfo
  {author} {\bibfnamefont {W.}~\bibnamefont {Schleich}},\ }\href@noop {} {\emph
  {\bibinfo {title} {{Mathematical Analysis of Evolution, Information, and
  Complexity, Weyl's Law}}}}\ (\bibinfo  {publisher} {Wiley-VCH, Weinheim,
  Germany},\ \bibinfo {year} {2009})\ pp.\ \bibinfo {pages} {1--71}\BibitemShut
  {NoStop}%
\bibitem [{\citenamefont {Savo}\ \emph {et~al.}(2017)\citenamefont {Savo},
  \citenamefont {Pierrat}, \citenamefont {Najar}, \citenamefont {Carminati},
  \citenamefont {Rotter},\ and\ \citenamefont {Gigan}}]{Savo2017}%
  \BibitemOpen
  \bibfield  {author} {\bibinfo {author} {\bibfnamefont {R.}~\bibnamefont
  {Savo}}, \bibinfo {author} {\bibfnamefont {R.}~\bibnamefont {Pierrat}},
  \bibinfo {author} {\bibfnamefont {U.}~\bibnamefont {Najar}}, \bibinfo
  {author} {\bibfnamefont {R.}~\bibnamefont {Carminati}}, \bibinfo {author}
  {\bibfnamefont {S.}~\bibnamefont {Rotter}}, \ and\ \bibinfo {author}
  {\bibfnamefont {S.}~\bibnamefont {Gigan}},\ }\href@noop {} {\bibfield
  {journal} {\bibinfo  {journal} {Science}\ }\textbf {\bibinfo {volume}
  {358}},\ \bibinfo {pages} {765} (\bibinfo {year} {2017})}\BibitemShut
  {NoStop}%
\bibitem [{\citenamefont {Mirlin}(2000)}]{Mirlin2000}%
  \BibitemOpen
  \bibfield  {author} {\bibinfo {author} {\bibfnamefont {A.}~\bibnamefont
  {Mirlin}},\ }\href@noop {} {\bibfield  {journal} {\bibinfo  {journal} {Phys.
  Rep.}\ }\textbf {\bibinfo {volume} {326}},\ \bibinfo {pages} {259} (\bibinfo
  {year} {2000})}\BibitemShut {NoStop}%
\bibitem [{\citenamefont {Lagendijk}\ \emph {et~al.}(2009)\citenamefont
  {Lagendijk}, \citenamefont {Tiggelen},\ and\ \citenamefont
  {Wiersma}}]{Lagendijk2009}%
  \BibitemOpen
  \bibfield  {author} {\bibinfo {author} {\bibfnamefont {A.}~\bibnamefont
  {Lagendijk}}, \bibinfo {author} {\bibfnamefont {B.}~\bibnamefont {Tiggelen}},
  \ and\ \bibinfo {author} {\bibfnamefont {D.}~\bibnamefont {Wiersma}},\
  }\href@noop {} {\bibfield  {journal} {\bibinfo  {journal} {Physics today}\
  }\textbf {\bibinfo {volume} {62}},\ \bibinfo {pages} {24} (\bibinfo {year}
  {2009})}\BibitemShut {NoStop}%
\bibitem [{\citenamefont {Yablonovitch}(1993)}]{Yablonovitch1993}%
  \BibitemOpen
  \bibfield  {author} {\bibinfo {author} {\bibfnamefont {E.}~\bibnamefont
  {Yablonovitch}},\ }\href {\doibase 10.1364/JOSAB.10.000283} {\bibfield
  {journal} {\bibinfo  {journal} {J. Opt. Soc. Am. B}\ }\textbf {\bibinfo
  {volume} {10}},\ \bibinfo {pages} {283} (\bibinfo {year} {1993})}\BibitemShut
  {NoStop}%
\bibitem [{\citenamefont {Yablonovitch}(1982)}]{Yablonovitch1982}%
  \BibitemOpen
  \bibfield  {author} {\bibinfo {author} {\bibfnamefont {E.}~\bibnamefont
  {Yablonovitch}},\ }\href@noop {} {\bibfield  {journal} {\bibinfo  {journal}
  {J. Opt. Soc. of Am.}\ }\textbf {\bibinfo {volume} {72}},\ \bibinfo {pages}
  {899} (\bibinfo {year} {1982})}\BibitemShut {NoStop}%
\bibitem [{\citenamefont {Chabanov}\ \emph {et~al.}(2000)\citenamefont
  {Chabanov}, \citenamefont {Stoytchev},\ and\ \citenamefont
  {Genack}}]{Chabanov2000}%
  \BibitemOpen
  \bibfield  {author} {\bibinfo {author} {\bibfnamefont {A.~A.}\ \bibnamefont
  {Chabanov}}, \bibinfo {author} {\bibfnamefont {M.}~\bibnamefont {Stoytchev}},
  \ and\ \bibinfo {author} {\bibfnamefont {A.~Z.}\ \bibnamefont {Genack}},\
  }\href {\doibase 10.1038/35009055} {\bibfield  {journal} {\bibinfo  {journal}
  {Nature}\ }\textbf {\bibinfo {volume} {404}},\ \bibinfo {pages} {850}
  (\bibinfo {year} {2000})}\BibitemShut {NoStop}%
\bibitem [{\citenamefont {Davy}\ and\ \citenamefont {Genack}(2018)}]{Davy2018}%
  \BibitemOpen
  \bibfield  {author} {\bibinfo {author} {\bibfnamefont {M.}~\bibnamefont
  {Davy}}\ and\ \bibinfo {author} {\bibfnamefont {A.~Z.}\ \bibnamefont
  {Genack}},\ }\href@noop {} {\bibfield  {journal} {\bibinfo  {journal} {Nat.
  Comm.}\ }\textbf {\bibinfo {volume} {9}},\ \bibinfo {pages} {4714} (\bibinfo
  {year} {2018})}\BibitemShut {NoStop}%
\bibitem [{\citenamefont {Abrahams}\ \emph {et~al.}(1979)\citenamefont
  {Abrahams}, \citenamefont {Anderson}, \citenamefont {Licciardello},\ and\
  \citenamefont {Ramakrishnan}}]{Abrahams1979}%
  \BibitemOpen
  \bibfield  {author} {\bibinfo {author} {\bibfnamefont {E.}~\bibnamefont
  {Abrahams}}, \bibinfo {author} {\bibfnamefont {P.~W.}\ \bibnamefont
  {Anderson}}, \bibinfo {author} {\bibfnamefont {D.~C.}\ \bibnamefont
  {Licciardello}}, \ and\ \bibinfo {author} {\bibfnamefont {T.~V.}\
  \bibnamefont {Ramakrishnan}},\ }\href@noop {} {\bibfield  {journal} {\bibinfo
   {journal} {Phys. Rev. Lett.}\ }\textbf {\bibinfo {volume} {42}},\ \bibinfo
  {pages} {673} (\bibinfo {year} {1979})}\BibitemShut {NoStop}%
\bibitem [{\citenamefont {Wigner}(1955)}]{Wigner1955}%
  \BibitemOpen
  \bibfield  {author} {\bibinfo {author} {\bibfnamefont {E.~P.}\ \bibnamefont
  {Wigner}},\ }\href@noop {} {\bibfield  {journal} {\bibinfo  {journal} {Phys.
  Rev.}\ }\textbf {\bibinfo {volume} {98}},\ \bibinfo {pages} {145} (\bibinfo
  {year} {1955})}\BibitemShut {NoStop}%
\bibitem [{\citenamefont {Smith}(1960)}]{Smith1960}%
  \BibitemOpen
  \bibfield  {author} {\bibinfo {author} {\bibfnamefont {F.~T.}\ \bibnamefont
  {Smith}},\ }\href@noop {} {\bibfield  {journal} {\bibinfo  {journal} {Phys.
  Rev.}\ }\textbf {\bibinfo {volume} {118}},\ \bibinfo {pages} {349} (\bibinfo
  {year} {1960})}\BibitemShut {NoStop}%
\bibitem [{\citenamefont {Kottos}(2005)}]{Kottos2005}%
  \BibitemOpen
  \bibfield  {author} {\bibinfo {author} {\bibfnamefont {T.}~\bibnamefont
  {Kottos}},\ }\href@noop {} {\bibfield  {journal} {\bibinfo  {journal} {J.
  Phys. A: Math. Gen.}\ }\textbf {\bibinfo {volume} {38}},\ \bibinfo {pages}
  {10761} (\bibinfo {year} {2005})}\BibitemShut {NoStop}%
\bibitem [{\citenamefont {Rotter}\ \emph {et~al.}(2011)\citenamefont {Rotter},
  \citenamefont {Ambichl},\ and\ \citenamefont {Libisch}}]{Rotter2011}%
  \BibitemOpen
  \bibfield  {author} {\bibinfo {author} {\bibfnamefont {S.}~\bibnamefont
  {Rotter}}, \bibinfo {author} {\bibfnamefont {P.}~\bibnamefont {Ambichl}}, \
  and\ \bibinfo {author} {\bibfnamefont {F.}~\bibnamefont {Libisch}},\ }\href
  {\doibase 10.1103/PhysRevLett.106.120602} {\bibfield  {journal} {\bibinfo
  {journal} {Phys. Rev. Lett.}\ }\textbf {\bibinfo {volume} {106}},\ \bibinfo
  {pages} {120602} (\bibinfo {year} {2011})}\BibitemShut {NoStop}%
\bibitem [{\citenamefont {G\'erardin}\ \emph {et~al.}(2016)\citenamefont
  {G\'erardin}, \citenamefont {Laurent}, \citenamefont {Ambichl}, \citenamefont
  {Prada}, \citenamefont {Rotter},\ and\ \citenamefont {Aubry}}]{Gerardin2016}%
  \BibitemOpen
  \bibfield  {author} {\bibinfo {author} {\bibfnamefont {B.}~\bibnamefont
  {G\'erardin}}, \bibinfo {author} {\bibfnamefont {J.}~\bibnamefont {Laurent}},
  \bibinfo {author} {\bibfnamefont {P.}~\bibnamefont {Ambichl}}, \bibinfo
  {author} {\bibfnamefont {C.}~\bibnamefont {Prada}}, \bibinfo {author}
  {\bibfnamefont {S.}~\bibnamefont {Rotter}}, \ and\ \bibinfo {author}
  {\bibfnamefont {A.}~\bibnamefont {Aubry}},\ }\href {\doibase
  10.1103/PhysRevB.94.014209} {\bibfield  {journal} {\bibinfo  {journal} {Phys.
  Rev. B}\ }\textbf {\bibinfo {volume} {94}},\ \bibinfo {pages} {014209}
  (\bibinfo {year} {2016})}\BibitemShut {NoStop}%
\bibitem [{\citenamefont {B\"ohm}\ \emph {et~al.}(2018)\citenamefont {B\"ohm},
  \citenamefont {Brandst\"otter}, \citenamefont {Ambichl}, \citenamefont
  {Rotter},\ and\ \citenamefont {Kuhl}}]{Boehm2018}%
  \BibitemOpen
  \bibfield  {author} {\bibinfo {author} {\bibfnamefont {J.}~\bibnamefont
  {B\"ohm}}, \bibinfo {author} {\bibfnamefont {A.}~\bibnamefont
  {Brandst\"otter}}, \bibinfo {author} {\bibfnamefont {P.}~\bibnamefont
  {Ambichl}}, \bibinfo {author} {\bibfnamefont {S.}~\bibnamefont {Rotter}}, \
  and\ \bibinfo {author} {\bibfnamefont {U.}~\bibnamefont {Kuhl}},\ }\href
  {\doibase 10.1103/PhysRevA.97.021801} {\bibfield  {journal} {\bibinfo
  {journal} {Phys. Rev. A}\ }\textbf {\bibinfo {volume} {97}},\ \bibinfo
  {pages} {021801} (\bibinfo {year} {2018})}\BibitemShut {NoStop}%
\bibitem [{\citenamefont {Xiong}\ \emph {et~al.}(2016)\citenamefont {Xiong},
  \citenamefont {Ambichl}, \citenamefont {Bromberg}, \citenamefont {Redding},
  \citenamefont {Rotter},\ and\ \citenamefont {Cao}}]{Xiong2016}%
  \BibitemOpen
  \bibfield  {author} {\bibinfo {author} {\bibfnamefont {W.}~\bibnamefont
  {Xiong}}, \bibinfo {author} {\bibfnamefont {P.}~\bibnamefont {Ambichl}},
  \bibinfo {author} {\bibfnamefont {Y.}~\bibnamefont {Bromberg}}, \bibinfo
  {author} {\bibfnamefont {B.}~\bibnamefont {Redding}}, \bibinfo {author}
  {\bibfnamefont {S.}~\bibnamefont {Rotter}}, \ and\ \bibinfo {author}
  {\bibfnamefont {H.}~\bibnamefont {Cao}},\ }\href {\doibase
  10.1103/PhysRevLett.117.053901} {\bibfield  {journal} {\bibinfo  {journal}
  {Phys. Rev. Lett.}\ }\textbf {\bibinfo {volume} {117}},\ \bibinfo {pages}
  {053901} (\bibinfo {year} {2016})}\BibitemShut {NoStop}%
\bibitem [{\citenamefont {Ambichl}\ \emph {et~al.}(2017)\citenamefont
  {Ambichl}, \citenamefont {Xiong}, \citenamefont {Bromberg}, \citenamefont
  {Redding}, \citenamefont {Cao},\ and\ \citenamefont {Rotter}}]{Ambichl2017}%
  \BibitemOpen
  \bibfield  {author} {\bibinfo {author} {\bibfnamefont {P.}~\bibnamefont
  {Ambichl}}, \bibinfo {author} {\bibfnamefont {W.}~\bibnamefont {Xiong}},
  \bibinfo {author} {\bibfnamefont {Y.}~\bibnamefont {Bromberg}}, \bibinfo
  {author} {\bibfnamefont {B.}~\bibnamefont {Redding}}, \bibinfo {author}
  {\bibfnamefont {H.}~\bibnamefont {Cao}}, \ and\ \bibinfo {author}
  {\bibfnamefont {S.}~\bibnamefont {Rotter}},\ }\href {\doibase
  10.1103/PhysRevX.7.041053} {\bibfield  {journal} {\bibinfo  {journal} {Phys.
  Rev. X}\ }\textbf {\bibinfo {volume} {7}},\ \bibinfo {pages} {041053}
  (\bibinfo {year} {2017})}\BibitemShut {NoStop}%
\bibitem [{\citenamefont {Carpenter}\ \emph {et~al.}(2015)\citenamefont
  {Carpenter}, \citenamefont {Eggleton},\ and\ \citenamefont
  {Schr{\"o}der}}]{Carpenter2015}%
  \BibitemOpen
  \bibfield  {author} {\bibinfo {author} {\bibfnamefont {J.}~\bibnamefont
  {Carpenter}}, \bibinfo {author} {\bibfnamefont {B.~J.}\ \bibnamefont
  {Eggleton}}, \ and\ \bibinfo {author} {\bibfnamefont {J.}~\bibnamefont
  {Schr{\"o}der}},\ }\href {https://doi.org/10.1038/nphoton.2015.188}
  {\bibfield  {journal} {\bibinfo  {journal} {Nature Photonics}\ }\textbf
  {\bibinfo {volume} {9}},\ \bibinfo {pages} {751} (\bibinfo {year} {2015})},\
  \bibinfo {note} {article}\BibitemShut {NoStop}%
\bibitem [{\citenamefont {Brandbyge}\ and\ \citenamefont
  {Tsukada}(1998)}]{Brandbyge1998}%
  \BibitemOpen
  \bibfield  {author} {\bibinfo {author} {\bibfnamefont {M.}~\bibnamefont
  {Brandbyge}}\ and\ \bibinfo {author} {\bibfnamefont {M.}~\bibnamefont
  {Tsukada}},\ }\href@noop {} {\bibfield  {journal} {\bibinfo  {journal} {Phys.
  Rev. B}\ }\textbf {\bibinfo {volume} {57}},\ \bibinfo {pages} {R15088}
  (\bibinfo {year} {1998})}\BibitemShut {NoStop}%
\bibitem [{\citenamefont {Davy}\ \emph {et~al.}(2015)\citenamefont {Davy},
  \citenamefont {Shi}, \citenamefont {Wang}, \citenamefont {Cheng},\ and\
  \citenamefont {Genack}}]{Davy2015}%
  \BibitemOpen
  \bibfield  {author} {\bibinfo {author} {\bibfnamefont {M.}~\bibnamefont
  {Davy}}, \bibinfo {author} {\bibfnamefont {Z.}~\bibnamefont {Shi}}, \bibinfo
  {author} {\bibfnamefont {J.}~\bibnamefont {Wang}}, \bibinfo {author}
  {\bibfnamefont {X.}~\bibnamefont {Cheng}}, \ and\ \bibinfo {author}
  {\bibfnamefont {A.~Z.}\ \bibnamefont {Genack}},\ }\href@noop {} {\bibfield
  {journal} {\bibinfo  {journal} {Phys. Rev. Lett.}\ }\textbf {\bibinfo
  {volume} {114}},\ \bibinfo {pages} {033901} (\bibinfo {year}
  {2015})}\BibitemShut {NoStop}%
\bibitem [{\citenamefont {Avishai}\ and\ \citenamefont
  {Band}(1985)}]{Avishai1985}%
  \BibitemOpen
  \bibfield  {author} {\bibinfo {author} {\bibfnamefont {Y.}~\bibnamefont
  {Avishai}}\ and\ \bibinfo {author} {\bibfnamefont {Y.~B.}\ \bibnamefont
  {Band}},\ }\href@noop {} {\bibfield  {journal} {\bibinfo  {journal} {Phys.
  Rev. B}\ }\textbf {\bibinfo {volume} {32}},\ \bibinfo {pages} {2674}
  (\bibinfo {year} {1985})}\BibitemShut {NoStop}%
\bibitem [{\citenamefont {Barnett}\ and\ \citenamefont
  {Loudon}(1996)}]{Barnett1996}%
  \BibitemOpen
  \bibfield  {author} {\bibinfo {author} {\bibfnamefont {S.~M.}\ \bibnamefont
  {Barnett}}\ and\ \bibinfo {author} {\bibfnamefont {R.}~\bibnamefont
  {Loudon}},\ }\href@noop {} {\bibfield  {journal} {\bibinfo  {journal} {Phys.
  Rev. Lett.}\ }\textbf {\bibinfo {volume} {77}},\ \bibinfo {pages} {2444}
  (\bibinfo {year} {1996})}\BibitemShut {NoStop}%
\bibitem [{\citenamefont {Fyodorov}\ \emph {et~al.}(2005)\citenamefont
  {Fyodorov}, \citenamefont {Savin},\ and\ \citenamefont
  {Sommers}}]{Fyodorov2005}%
  \BibitemOpen
  \bibfield  {author} {\bibinfo {author} {\bibfnamefont {Y.~V.}\ \bibnamefont
  {Fyodorov}}, \bibinfo {author} {\bibfnamefont {D.}~\bibnamefont {Savin}}, \
  and\ \bibinfo {author} {\bibfnamefont {H.}~\bibnamefont {Sommers}},\
  }\href@noop {} {\bibfield  {journal} {\bibinfo  {journal} {J. Phys. A Math.
  Theor.}\ }\textbf {\bibinfo {volume} {38}},\ \bibinfo {pages} {10731}
  (\bibinfo {year} {2005})}\BibitemShut {NoStop}%
\bibitem [{\citenamefont {Liew}\ \emph {et~al.}(2014)\citenamefont {Liew},
  \citenamefont {Popoff}, \citenamefont {Mosk}, \citenamefont {Vos},\ and\
  \citenamefont {Cao}}]{Liew2014}%
  \BibitemOpen
  \bibfield  {author} {\bibinfo {author} {\bibfnamefont {S.~F.}\ \bibnamefont
  {Liew}}, \bibinfo {author} {\bibfnamefont {S.~M.}\ \bibnamefont {Popoff}},
  \bibinfo {author} {\bibfnamefont {A.~P.}\ \bibnamefont {Mosk}}, \bibinfo
  {author} {\bibfnamefont {W.~L.}\ \bibnamefont {Vos}}, \ and\ \bibinfo
  {author} {\bibfnamefont {H.}~\bibnamefont {Cao}},\ }\href {\doibase
  10.1103/PhysRevB.89.224202} {\bibfield  {journal} {\bibinfo  {journal} {Phys.
  Rev. B}\ }\textbf {\bibinfo {volume} {89}},\ \bibinfo {pages} {224202}
  (\bibinfo {year} {2014})}\BibitemShut {NoStop}%
\bibitem [{\citenamefont {Ambichl}(2016)}]{AmbichlPhd2016}%
  \BibitemOpen
  \bibfield  {author} {\bibinfo {author} {\bibfnamefont {P.}~\bibnamefont
  {Ambichl}},\ }\emph {\bibinfo {title} {Coherent Wave Transport: Time Delay
  and Beyond}},\ \href@noop {} {Ph.D. thesis},\ \bibinfo  {school} {Vienna
  University of Technology, Institute for Theoretical Physics} (\bibinfo {year}
  {2016})\BibitemShut {NoStop}%
\bibitem [{\citenamefont {Durand}\ \emph {et~al.}(2019)\citenamefont {Durand},
  \citenamefont {Popoff}, \citenamefont {Carminati},\ and\ \citenamefont
  {Goetschy}}]{Durand2019}%
  \BibitemOpen
  \bibfield  {author} {\bibinfo {author} {\bibfnamefont {M.}~\bibnamefont
  {Durand}}, \bibinfo {author} {\bibfnamefont {S.~M.}\ \bibnamefont {Popoff}},
  \bibinfo {author} {\bibfnamefont {R.}~\bibnamefont {Carminati}}, \ and\
  \bibinfo {author} {\bibfnamefont {A.}~\bibnamefont {Goetschy}},\ }\href@noop
  {} {\bibfield  {journal} {\bibinfo  {journal} {Phys. Rev. Lett.}\ }\textbf
  {\bibinfo {volume} {123}},\ \bibinfo {pages} {243901} (\bibinfo {year}
  {2019})}\BibitemShut {NoStop}%
\bibitem [{\citenamefont {Carminati}\ and\ \citenamefont
  {Sáenz}(2009)}]{Carminati2009}%
  \BibitemOpen
  \bibfield  {author} {\bibinfo {author} {\bibfnamefont {R.}~\bibnamefont
  {Carminati}}\ and\ \bibinfo {author} {\bibfnamefont {J.}~\bibnamefont
  {Sáenz}},\ }\href@noop {} {\bibfield  {journal} {\bibinfo  {journal}
  {Physical Review Letters}\ }\textbf {\bibinfo {volume} {102}},\ \bibinfo
  {pages} {093902} (\bibinfo {year} {2009})}\BibitemShut {NoStop}%
\bibitem [{NGS()}]{NGSolve}%
  \BibitemOpen
  \href@noop {} {}\bibinfo {note} {{Netgen/NGSolve} multiphysics finite element
  software, https://ngsolve.org}\BibitemShut {NoStop}%
\bibitem [{\citenamefont {Sch{\"o}berl}(1997)}]{Schöberl1997}%
  \BibitemOpen
  \bibfield  {author} {\bibinfo {author} {\bibfnamefont {J.}~\bibnamefont
  {Sch{\"o}berl}},\ }\href {\doibase 10.1007/s007910050004} {\bibfield
  {journal} {\bibinfo  {journal} {Computing and Visualization in Science}\
  }\textbf {\bibinfo {volume} {1}},\ \bibinfo {pages} {41} (\bibinfo {year}
  {1997})}\BibitemShut {NoStop}%
\bibitem [{\citenamefont {Sch{\"o}berl}(2014)}]{Schöberl2014}%
  \BibitemOpen
  \bibfield  {author} {\bibinfo {author} {\bibfnamefont {J.}~\bibnamefont
  {Sch{\"o}berl}},\ }\href@noop {} {\enquote {\bibinfo {title} {{C++11
  Implementation of Finite Elements in NGSolve}},}\ }\bibinfo {howpublished}
  {ASC Report, Institute for Analysis and Scientific Computing, Vienna
  University of Technology} (\bibinfo {year} {2014})\BibitemShut {NoStop}%
\bibitem [{\citenamefont {Brandst{\"o}tter}\ \emph {et~al.}(2019)\citenamefont
  {Brandst{\"o}tter}, \citenamefont {Girschik}, \citenamefont {Ambichl},\ and\
  \citenamefont {Rotter}}]{Brandstoetter2019}%
  \BibitemOpen
  \bibfield  {author} {\bibinfo {author} {\bibfnamefont {A.}~\bibnamefont
  {Brandst{\"o}tter}}, \bibinfo {author} {\bibfnamefont {A.}~\bibnamefont
  {Girschik}}, \bibinfo {author} {\bibfnamefont {P.}~\bibnamefont {Ambichl}}, \
  and\ \bibinfo {author} {\bibfnamefont {S.}~\bibnamefont {Rotter}},\ }\href
  {\doibase 10.1073/pnas.1905217116} {\bibfield  {journal} {\bibinfo  {journal}
  {Proceedings of the National Academy of Sciences}\ }\textbf {\bibinfo
  {volume} {116}},\ \bibinfo {pages} {13260} (\bibinfo {year}
  {2019})}\BibitemShut {NoStop}%
\end{thebibliography}

\begin{thebibliography}{8}%
\makeatletter
\providecommand \@ifxundefined [1]{%
 \@ifx{#1\undefined}
}%
\providecommand \@ifnum [1]{%
 \ifnum #1\expandafter \@firstoftwo
 \else \expandafter \@secondoftwo
 \fi
}%
\providecommand \@ifx [1]{%
 \ifx #1\expandafter \@firstoftwo
 \else \expandafter \@secondoftwo
 \fi
}%
\providecommand \natexlab [1]{#1}%
\providecommand \enquote  [1]{``#1''}%
\providecommand \bibnamefont  [1]{#1}%
\providecommand \bibfnamefont [1]{#1}%
\providecommand \citenamefont [1]{#1}%
\providecommand \href@noop [0]{\@secondoftwo}%
\providecommand \href [0]{\begingroup \@sanitize@url \@href}%
\providecommand \@href[1]{\@@startlink{#1}\@@href}%
\providecommand \@@href[1]{\endgroup#1\@@endlink}%
\providecommand \@sanitize@url [0]{\catcode `\\12\catcode `\$12\catcode
  `\&12\catcode `\#12\catcode `\^12\catcode `\_12\catcode `\%12\relax}%
\providecommand \@@startlink[1]{}%
\providecommand \@@endlink[0]{}%
\providecommand \url  [0]{\begingroup\@sanitize@url \@url }%
\providecommand \@url [1]{\endgroup\@href {#1}{\urlprefix }}%
\providecommand \urlprefix  [0]{URL }%
\providecommand \Eprint [0]{\href }%
\providecommand \doibase [0]{http://dx.doi.org/}%
\providecommand \selectlanguage [0]{\@gobble}%
\providecommand \bibinfo  [0]{\@secondoftwo}%
\providecommand \bibfield  [0]{\@secondoftwo}%
\providecommand \translation [1]{[#1]}%
\providecommand \BibitemOpen [0]{}%
\providecommand \bibitemStop [0]{}%
\providecommand \bibitemNoStop [0]{.\EOS\space}%
\providecommand \EOS [0]{\spacefactor3000\relax}%
\providecommand \BibitemShut  [1]{\csname bibitem#1\endcsname}%
\let\auto@bib@innerbib\@empty
\bibitem [{\citenamefont {Schwinger}(1951)}]{Schwinger1951}%
  \BibitemOpen
  \bibfield  {author} {\bibinfo {author} {\bibfnamefont {J.}~\bibnamefont
  {Schwinger}},\ }\href@noop {} {\bibfield  {journal} {\bibinfo  {journal}
  {Phys. Rev.}\ }\textbf {\bibinfo {volume} {82}},\ \bibinfo {pages} {664}
  (\bibinfo {year} {1951})}\BibitemShut {NoStop}%
\bibitem [{\citenamefont {Krein}(1962)}]{Krein1962}%
  \BibitemOpen
  \bibfield  {author} {\bibinfo {author} {\bibfnamefont {M.~G.}\ \bibnamefont
  {Krein}},\ }\href@noop {} {\bibfield  {journal} {\bibinfo  {journal} {Dokl.
  Akad. Nauk SSSR}\ }\textbf {\bibinfo {volume} {144}},\ \bibinfo {pages} {475}
  (\bibinfo {year} {1962})}\BibitemShut {NoStop}%
\bibitem [{\citenamefont {Birman}\ and\ \citenamefont
  {Yafaev}(1992)}]{Birman1992}%
  \BibitemOpen
  \bibfield  {author} {\bibinfo {author} {\bibfnamefont {M.~S.}\ \bibnamefont
  {Birman}}\ and\ \bibinfo {author} {\bibfnamefont {D.~R.}\ \bibnamefont
  {Yafaev}},\ }\href@noop {} {\bibfield  {journal} {\bibinfo  {journal}
  {Algebra i Analiz}\ }\textbf {\bibinfo {volume} {4}},\ \bibinfo {pages} {1}
  (\bibinfo {year} {1992})}\BibitemShut {NoStop}%
\bibitem [{\citenamefont {Iannaccone}(1995)}]{Iannaccone1995}%
  \BibitemOpen
  \bibfield  {author} {\bibinfo {author} {\bibfnamefont {G.}~\bibnamefont
  {Iannaccone}},\ }\href@noop {} {\bibfield  {journal} {\bibinfo  {journal}
  {Phys. Rev. B}\ }\textbf {\bibinfo {volume} {51}},\ \bibinfo {pages} {4727}
  (\bibinfo {year} {1995})}\BibitemShut {NoStop}%
\bibitem [{\citenamefont {Breit}\ and\ \citenamefont
  {Wigner}(1936)}]{Breit1936}%
  \BibitemOpen
  \bibfield  {author} {\bibinfo {author} {\bibfnamefont {G.}~\bibnamefont
  {Breit}}\ and\ \bibinfo {author} {\bibfnamefont {E.}~\bibnamefont {Wigner}},\
  }\href {\doibase 10.1103/PhysRev.49.519} {\bibfield  {journal} {\bibinfo
  {journal} {Phys. Rev.}\ }\textbf {\bibinfo {volume} {49}},\ \bibinfo {pages}
  {519} (\bibinfo {year} {1936})}\BibitemShut {NoStop}%
\bibitem [{\citenamefont {Davy}\ \emph {et~al.}(2012)\citenamefont {Davy},
  \citenamefont {Shi},\ and\ \citenamefont {Genack}}]{Davy2012}%
  \BibitemOpen
  \bibfield  {author} {\bibinfo {author} {\bibfnamefont {M.}~\bibnamefont
  {Davy}}, \bibinfo {author} {\bibfnamefont {Z.}~\bibnamefont {Shi}}, \ and\
  \bibinfo {author} {\bibfnamefont {A.~Z.}\ \bibnamefont {Genack}},\
  }\href@noop {} {\bibfield  {journal} {\bibinfo  {journal} {Phys. Rev. B}\
  }\textbf {\bibinfo {volume} {85}},\ \bibinfo {pages} {035105} (\bibinfo
  {year} {2012})}\BibitemShut {NoStop}%
\bibitem [{\citenamefont {Leseur}\ \emph {et~al.}(2014)\citenamefont {Leseur},
  \citenamefont {Pierrat}, \citenamefont {S{\'a}enz},\ and\ \citenamefont
  {Carminati}}]{Leseur2014}%
  \BibitemOpen
  \bibfield  {author} {\bibinfo {author} {\bibfnamefont {O.}~\bibnamefont
  {Leseur}}, \bibinfo {author} {\bibfnamefont {R.}~\bibnamefont {Pierrat}},
  \bibinfo {author} {\bibfnamefont {J.~J.}\ \bibnamefont {S{\'a}enz}}, \ and\
  \bibinfo {author} {\bibfnamefont {R.}~\bibnamefont {Carminati}},\ }\href@noop
  {} {\bibfield  {journal} {\bibinfo  {journal} {Phys. Rev. A}\ }\textbf
  {\bibinfo {volume} {90}},\ \bibinfo {pages} {053827} (\bibinfo {year}
  {2014})}\BibitemShut {NoStop}%
\bibitem [{\citenamefont {Davy}\ and\ \citenamefont {Genack}(2018)}]{Davy2018}%
  \BibitemOpen
  \bibfield  {author} {\bibinfo {author} {\bibfnamefont {M.}~\bibnamefont
  {Davy}}\ and\ \bibinfo {author} {\bibfnamefont {A.~Z.}\ \bibnamefont
  {Genack}},\ }\href@noop {} {\bibfield  {journal} {\bibinfo  {journal} {Nat.
  Comm.}\ }\textbf {\bibinfo {volume} {9}},\ \bibinfo {pages} {4714} (\bibinfo
  {year} {2018})}\BibitemShut {NoStop}%
\end{thebibliography}

\providecommand{\noopsort}[1]{}\providecommand{\singleletter}[1]{#1}%

\end{document}